\documentclass[review]{elsarticle}

\usepackage{lineno}

\usepackage{setspace}
\usepackage{subfigure}
\usepackage{courier}
\usepackage{amsfonts}
\usepackage{amsmath,amssymb,mathtools}
\usepackage[ruled]{algorithm2e}
\usepackage{graphics}
\usepackage{url}
\usepackage{bm}
\usepackage[below]{placeins}
\usepackage{booktabs}
\usepackage[table,xcdraw]{xcolor}
\usepackage{multirow}
\usepackage{graphicx}
\usepackage[normalem]{ulem}
\usepackage{pifont} % 特殊符号
\usepackage{enumerate} 
\usepackage{float}
\usepackage[backref]{hyperref}
\usepackage{lscape}
\DeclareGraphicsExtensions{.pdf,.jpeg,.png}
\modulolinenumbers[5]

\journal{Information Sciences}

\biboptions{sort&compress}

\begin{document}

\begin{frontmatter}

\title{New Framework for Code-Mapping-based Reversible Data Hiding in JPEG Images}

\author[mymainaddress,mysecondaryaddress]{Yang Du}
\ead{duyang@stu.ahu.edu.cn}
    
\author[mymainaddress,tutoraddress]{Zhaoxia Yin\corref{mycorrespondingauthor}}
\ead{zhaoxia.edu@gmail.com}
    
\cortext[mycorrespondingauthor]{Corresponding author}
    
\address[mymainaddress]{Key Lab of Intelligent Computing and Signal Processing of Ministry of Education, School of Computer Science and Technology, Anhui University, Hefei 230601, China}
\address[mysecondaryaddress]{Anhui Key Laboratory of Multimodal Cognitive Computation, School of Computer Science and Technology, Anhui University, Hefei 230601, China}
\address[tutoraddress]{School of Communication \& Electronic Engineering, East China Normal University, Shanghai 200241, China}

\begin{abstract}
    Code mapping (CM) is an efficient technique for reversible data hiding (RDH) in JPEG images, which embeds data by constructing a mapping relationship between the used and unused codes in the JPEG bitstream. This study presents a new framework for designing a CM-based RDH method. First, a new code mapping strategy is proposed to suppress file size expansion and improve applicability. Based on our proposed strategy, the mapped codes are redefined by creating a new Huffman table rather than selecting them from the unused codes in the original Huffman table. The critical issue of designing the CM-based RDH method, that is, constructing code mapping, is converted into a combinatorial optimization problem. This study proposes a novel CM-based RDH method that utilizes a genetic algorithm (GA). The experimental results demonstrate that the proposed method achieves a high embedding capacity with no signal distortion while suppressing file size expansion.

\end{abstract}


\end{frontmatter}

% \linenumbers

\section{Introduction} 

\par Reversible data hiding (RDH) is a technique that can imperceptibly embed additional data into media and then restore the media error-free. The past decades have witnessed the rapid development of RDH techniques in different application scenarios, such as remote sensing, medical image processing, and image authentication \cite{KHAN2014251}. Many RDH methods have been developed for uncompressed images, including lossless compression \cite{fridrich2002lossless_a}\cite{celik2005lossless}, difference expansion (DE) \cite{tian2003reversible}\cite{dragoi2014local}, histogram shifting (HS) \cite{ni2006reversible,li2013general,Hu2021}, and prediction error expansion (PEE) \cite{WANG201516,WENG2019136,KUMAR202096,WENG202113}. 

\par JPEG RDH research has attracted considerable attention in recent years because JPEG is the most commonly used image compression format in applications. JPEG RDH methods mainly focus on two aspects: modifying the quantized discrete cosine transform (DCT) coefficients \cite{fridrich2002lossless,fridrich2004lossless,lv2018novel,Ong2013,Long2016,xuan2007reversible,huang2016reversible,qian2017reversible,wedaj2017improved,hou2018reversible,he2020reversible,li2020reversible,yin2020reversible,xiao2020efficient,xiao2021reversible} and manipulating the JPEG bitstream \cite{qian2012lossless,hu2013improved,qiu2018lossless,zhang2020improved,qiu2020optimized,du2020gvm,zhang2020reversible}. The first type of methods, which focus on DCT coefficients are mainly based on lossless compression \cite{fridrich2002lossless,fridrich2004lossless,lv2018novel}, zero-run value (ZRV) pair rotation \cite{Ong2013,Long2016}, and HS \cite{xuan2007reversible,huang2016reversible,qian2017reversible,wedaj2017improved,hou2018reversible,he2020reversible,yin2020reversible,li2020reversible,xiao2020efficient,xiao2021reversible}. Because the DCT coefficients are modified directly during data embedding, these RDH methods cause file size expansion and image distortion. Among the aforementioned methods, HS-based methods \cite{xuan2007reversible,huang2016reversible,qian2017reversible,wedaj2017improved,hou2018reversible,he2020reversible,yin2020reversible,li2020reversible,xiao2020efficient,xiao2021reversible}, which focus on the modification of DCT coefficients to balance signal quality and file size preservation, have been the most widely studied. 

\par The second type of JPEG RDH methods is mainly based on code mapping (CM) \cite{qian2012lossless,hu2013improved,qiu2018lossless,zhang2020improved,zhang2020reversible,qiu2020optimized,du2020gvm}, which embeds data by building the mapping between used variable length codes (VLC) and unused VLCs in the JPEG bitstream. In \cite{mobasseri2010}, data were embedded into the bitstream by flipping one bit of the code. Mobasseri \emph{et~al.} applied an error concealment technique to minimize image quality distortion. Inspired by previous research \cite{mobasseri2010}, Qian and Zhang \cite{qian2012lossless} mapped used VLCs with several unused VLCs of identical lengths, guaranteeing no quality distortion and preserving the file size of the bitstream. Subsequently, a number of CM-based RDH methods \cite{hu2013improved,qiu2018lossless,zhang2020improved} aimed to improve the capacity while keeping the length of the marked JPEG bitstream unchanged. Several CM-based methods \cite{zhang2020reversible,qiu2020optimized,du2020gvm} have recently attempted to improve capacity by undermining file size increments. Qiu \emph{et~al.} \cite{qiu2020optimized} first employed a relay-transfer-based algorithm to preprocess the JPEG bitstream and then embedded data into the processed bitstream to improve its capacity. Du \emph{et~al.} \cite{du2020gvm} proposed the reordering of run-size values and established a simulated embedding model to find a better mapping relationship. Zhang \emph{et~al.} \cite{zhang2020reversible} utilized a value transfer matrix to simulate a theoretical model and designed optimization rules to build a mapping relationship.

\par In general, CM-based RDH studies \cite{qian2012lossless,hu2013improved,qiu2018lossless,zhang2020improved,zhang2020reversible,qiu2020optimized,du2020gvm} have two major advantages. On the one hand, CM-based RDH methods hide the secret bits directly into the bitstream without modifying the DCT coefficients; thus, no distortion is generated. Therefore, the image restoration process can be canceled after data extraction, because the image content remains unchanged after data embedding. On the other hand, it is more efficient because CM-based methods require neither full decompression nor recompression. Therefore, the CM technique is a better choice among existing approaches for JPEG RDH.

\par However, some problems in CM-based JPEG RDH research still exist. First, previous CM-based methods \cite{qian2012lossless,hu2013improved,qiu2018lossless,zhang2020improved,zhang2020reversible,qiu2020optimized,du2020gvm} rarely consider file size as an optimization goal during code mapping construction. However, file size preservation is considered a critical issue when designing a satisfactory JPEG RDH method.  Although recent CM-based studies \cite{zhang2020reversible,qiu2020optimized,du2020gvm} achieved high embedding capacities, the file size also increased as the capacity increased. Second, the aforementioned CM-based RDH methods \cite{qian2012lossless,hu2013improved,qiu2018lossless,zhang2020improved,zhang2020reversible,qiu2020optimized,du2020gvm} apply only to the bitstream encoded with the standard Huffman table (hereafter referred to as \emph{std-bitstream}). With the massive number of images growing on the Internet, the bitstream encoded with the optimized Huffman table (hereafter referred to as \emph{opt-bitstream}) is preferred over the std-bitstream because it requires less storage space. In contrast to the std-bitstream, all VLCs in the opt-bitstream are used; therefore, the code mapping cannot be constructed. This means that the aforementioned CM-based methods \cite{qian2012lossless,hu2013improved,qiu2018lossless,zhang2020improved,zhang2020reversible,du2020gvm,qiu2020optimized} are unavailable in the opt-bitstream.

\par In this study, we proposed a novel framework for designing a CM-based RDH method that achieves good file size preservation performance, and demonstrated it with a novel CM-based RDH method. The main contributions of this study are summarized as follows:

\begin{itemize}
    \item A new code mapping strategy is proposed to improve file size preservation and applicability performance.
    \item A new framework to design a CM-based method is provided. Under this framework, the critical issue of the CM-based method, i.e., code mapping construction, is converted into solving a combinatorial optimization problem with different given constraints.
    \item As a realization of the proposed framework, a novel CM-based JPEG RDH method is proposed by employing a genetic algorithm (GA) to solve the optimization problem.
    \item Comprehensive experiments demonstrate that the proposed method effectively suppresses the file size expansion compared to previous CM-based RDH methods and HS-based RDH methods.
\end{itemize}

\par The remainder of this paper is organized as follows. Section \ref{section2} provides an overview of the JPEG bitstream structure and the basic CM-based RDH method. Section \ref{section3} describes the proposed code-mapping strategy. Section \ref{section4} presents a detailed universal framework. Section \ref{section5} presents the new CM-based RDH method, Section \ref{section6} describes the experimental results, and Section \ref{section7} concludes the paper.

\section{Background}\label{section2}

\par In this section, we first introduce the baseline JPEG bitstream structure. We then briefly present the traditional CM-based RDH framework. 

\subsection{JPEG bitstream structure}

\par
\begin{figure}[!ht]
    \centering
    \includegraphics[width = 0.6\textwidth]{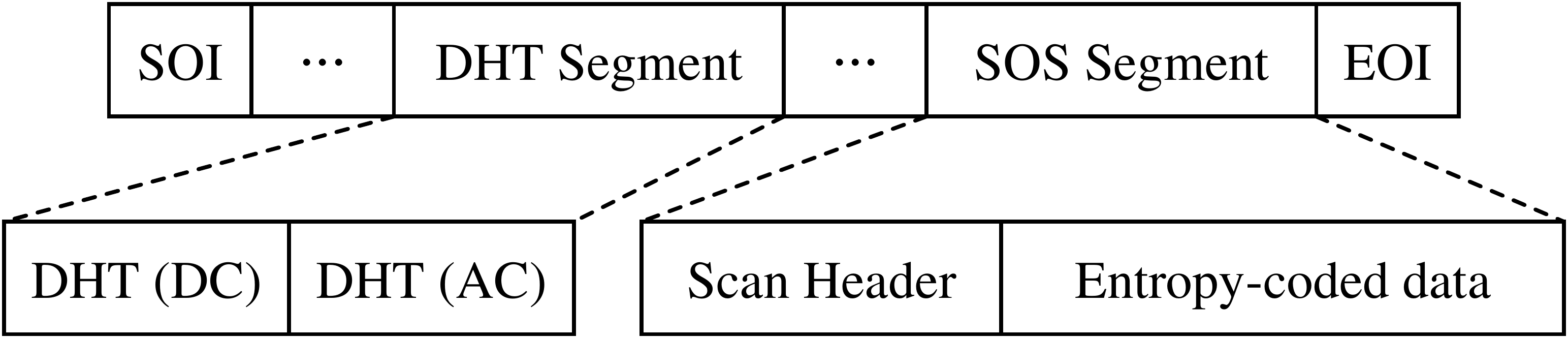}
    \caption{JPEG bitstream structure.}
    \label{fig_bitstream}
\end{figure}

 For brevity, we only introduce the syntax of the JPEG baseline for grayscale images. The JPEG bitstream consists of a file header and entropy-coded data, as shown in Fig. \ref{fig_bitstream}. A JPEG bitstream starts with the marker \emph{start-of-image} (SOI) and ends with the marker \emph{end-of-image} (EOI). There are several segments between SOI and EOI, including the define Huffman table (DHT) segment and the start-of-scan (SOS) segment related to the CM-based RDH.

\par
\begin{figure}[htbp]
    \centering
    \includegraphics[width = 0.6\textwidth]{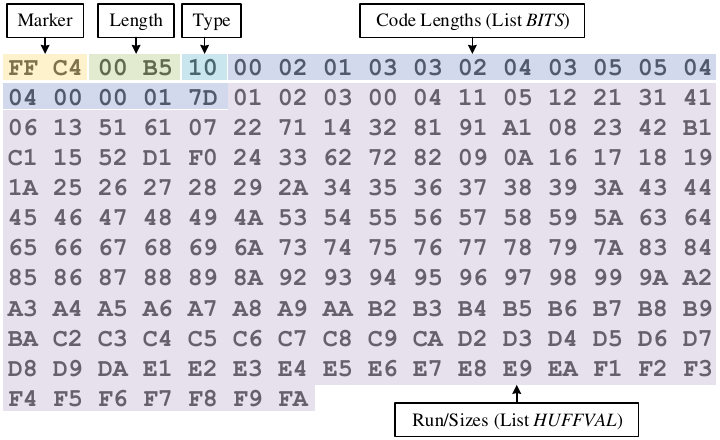}
    \caption{DHT Segment for AC coefficients contains the standard Huffman table information defined in the JPEG standard \cite{standard1992information}. \texttt{0xFFC4} is the marker that represents the DHT segment. \texttt{0x00B5} denotes the segment length except for the marker \texttt{0xFFC4}, which equals to B5$_{(16)}=181$ bytes. \texttt{0x10} specifies the type of DHT segment for AC coefficients. The lists BITS and HUFFVAL record the Huffman table information.}
    \label{fig_dht}
\end{figure}

 As described in the JPEG standard \cite{standard1992information}, the quantized AC coefficients are first compressed into the form of $(run/size, value)$. $value$ is the amplitude of the next non-zero AC coefficient, which is further encoded, and the generated codes are referred to as appended bits. Each symbol $run/size$ (RS) is encoded using a unique VLC. Fig. \ref{fig_dht} presents the DHT segment containing the RS information to construct the Huffman table. As shown in Fig. \ref{fig_dht}, the DHT segment mainly includes two lists: BITS and HUFFVAL. The BITS list contains VLCs of different lengths. For example, the first byte ``00'' means the code length of one is zero. The HUFFVAL list contains the hexadecimal forms of the RSs. For example, the first RS ``0/1'' is recorded as \texttt{0x01}. In a std-bitstream, the HUFFVAL list occupied 162 bytes, covering all 162 RSs. Each RS in the HUFFVAL list corresponds to one VLC defined in the standard Huffman table. However, the size of the HUFFVAL list in the opt-bitstream depends on nonzero-frequency RSs (NFRS). A Huffman table can be constructed based on these two lists. The SOS segment includes a scan header and entropy-coded data. The entropy-coded data include the compressed forms of the DC and AC coefficients, that is, the VLCs and appended bits, respectively.

\subsection{Traditional CM-based RDH framework}

\begin{figure*}[!ht]
    \centering
    \includegraphics[width = 0.8\textwidth]{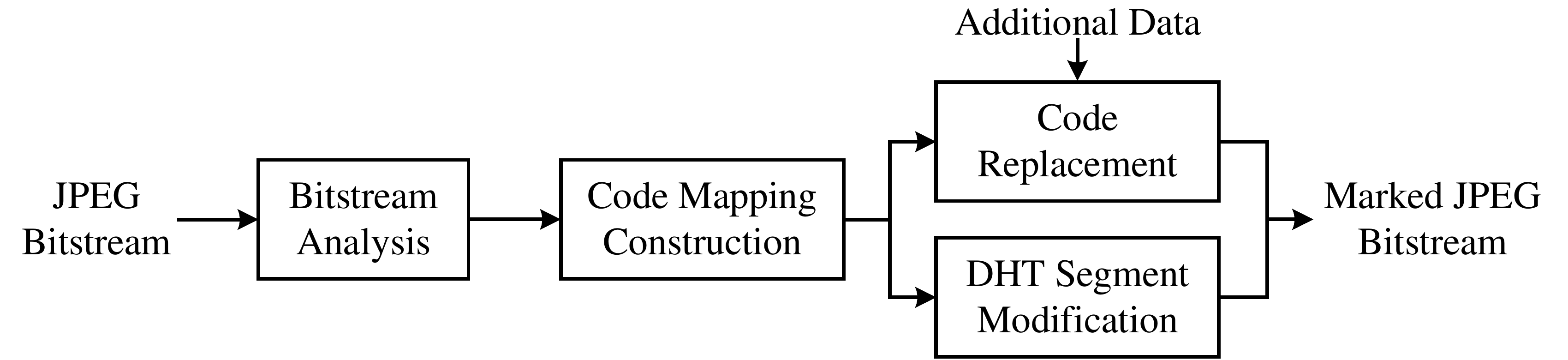}
    \caption{Traditional framework of CM-based JPEG RDH.}
    \label{fig_framework_traditional}
\end{figure*}

\par The traditional CM-based RDH framework is depicted in Fig. \ref{fig_framework_traditional}. In the bitstream analysis phase, the JPEG bitstream is parsed and all VLCs are constructed according to the DHT segment. Then, the occurrence of each VLC used in the entropy-coded data was calculated. In the code-mapping construction phase, some mapping rules are designed to construct code mappings between the used and unused VLCs. Early file-size-preserving CM-based methods \cite{qian2012lossless,hu2013improved,qiu2018lossless,zhang2020improved} mapped the used VLC to unused VLCs with the same code lengths to ensure that the file size of the bitstream remained invariant, causing limited embedding capacity. To improve the capacity, recent CM-based RDH research \cite{qian2012lossless,hu2013improved,qiu2018lossless,zhang2020improved} built a mapping among VLCs across different code lengths. After determining the code mapping relationship, additional data can be embedded by replacing the used VLCs with VLCs in the same mapping set. To maintain the decoded image pixels unchanged after embedding, the zero-frequency RS (ZFRS) of the corresponding VLC in the same mapping set is replaced by the NFRS in the DHT segment. 

\section{Proposed code mapping strategy}\label{section3}

\par During the code mapping construction, previous CM-based studies \cite{qian2012lossless,hu2013improved,qiu2018lossless,zhang2020improved,zhang2020reversible,du2020gvm,qiu2020optimized} directly select the required VLCs from existing unused VLCs and map them to used VLCs. We refer to this code-mapping strategy as the traditional code-mapping strategy. With the traditional mapping strategy, during the DHT segment modification, the selected ZFRSs are substituted by NFRSs to represent the mapping relationship, and the selected ZFRSs remain unchanged. This means that the Huffman table constructed using the modified DHT segment remains similar to the standard Huffman table. However, the standard Huffman table is defined by the average statistics obtained from a large collection of data, which does not exactly match the actual distribution of RSs in a JPEG bitstream; thus, coding redundancy exists. Recent CM-based RDH studies \cite{qian2012lossless,hu2013improved,qiu2018lossless,zhang2020improved} achieved a high embedding capacity by constructing mappings among VLCs across different code lengths. These studies used a traditional code-mapping strategy. Although these studies proposed some means to control file size expansion, the file size increment is still apparent.

\begin{figure*}[!ht]
    \centering
    \includegraphics[width=1\textwidth]{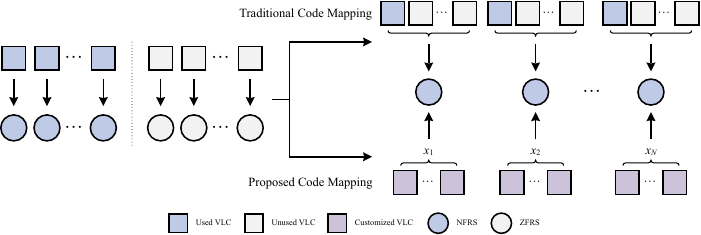}
    \caption{Difference between traditional code mapping strategy and our proposed code mapping strategy. In the proposed strategy, the required VLCs are all customized and mapped to NFRSs, rather than being selected from the standard Huffman table as in previous CM-based methods. Therefore, the proposed code mapping strategy applies to both the std-bitstream and opt-bitstream.}
    \label{fig_idea}
\end{figure*}

\par To further suppress the file size expansion, we propose a new code mapping strategy to guide the construction of the mapping relationship. The idea of the proposed strategy is to redefine the required VLCs, that is, all the VLCs used to construct the mapping relationship, by thoroughly customizing the Huffman table and mapping them to NFRSs. The difference between the traditional code mapping strategy and the proposed code mapping strategy is shown in Fig. \ref{fig_idea}. In the proposed strategy, each NFRS can be mapped using multiple VLCs. Instead of the Huffman table in the original JPEG bitstream, the required VLCs only depend on the distribution of NFRSs and the given constraints during the data embedding; therefore, the method designed based on the proposed strategy applies to both the opt-bitstream and std-bitstream. During the DHT segment modification, we added several copies of the original NFRSs to the HUFFVAL list, which is necessary for constructing the code mapping. Because only the necessary RSs are stored in the DHT segment, the Huffman table constructed using the customized DHT segment is similar to the optimized Huffman table. Therefore, coding redundancy can be significantly reduced; thus, the file size will be well preserved.

\section{New framework for CM-based RDH}\label{section4}

\begin{figure*}[!ht]
    \centering
    \includegraphics[width = 1\textwidth]{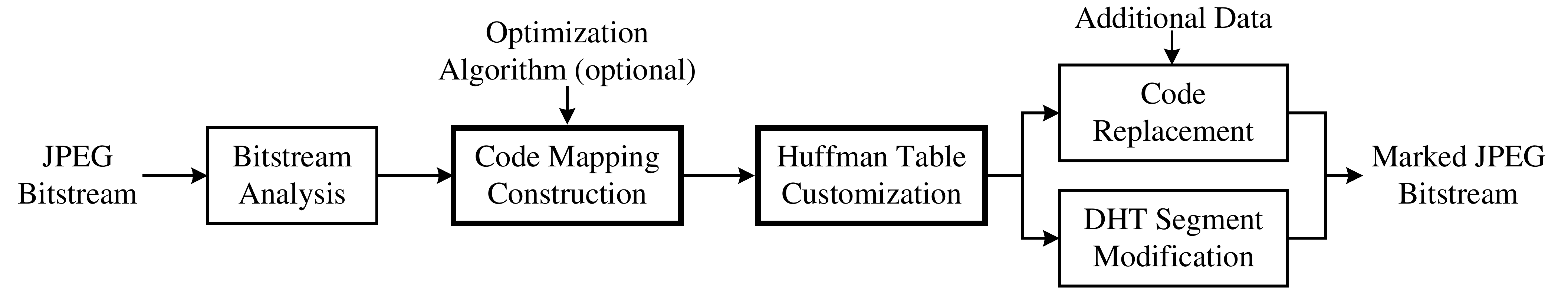}
    \caption{New framework of CM-based JPEG RDH.}
    \label{fig_framework_proposed}
\end{figure*}

\par The proposed framework for CM-based RDH is shown in Fig. \ref{fig_framework_proposed}. First, the original Huffman table was constructed by parsing the BITS and HUFFVAL lists in the DHT segment. Using the original Huffman table, the occurrences of the used VLCs in the entropy-coded data were calculated, and the statistical results of the RS frequency were obtained. According to the RS frequency and given constraints, the NFRSs were mapped with multiple VLCs adaptively to construct an optimal code mapping relationship. In the proposed framework, the code mapping construction is formulated as a combinatorial optimization problem and can be solved by leveraging optimization algorithms. We then constructed a customized Huffman table according to the optimal solution. The additional data were then embedded into the entropy-coded data by replacing the original VLCs and modifying the DHT segment. Finally, the marked JPEG bitstream was synthesized from the new entropy-coded data and new file header. The marked JPEG bitstream can still be decoded directly using the most widely used decoders.

\par Compared with the traditional framework of CM-based JPEG RDH, the proposed framework has two key issues. The first issue is to determine the number of VLCs mapped to each NFRS under the given constraints. To address this issue, we formulated the code-mapping construction as a combinatorial optimization problem, which can be solved using the optimization algorithm, as discussed in Section \ref{section4.1}. The second issue is the process of Huffman table customization, which is detailed in Section \ref{section4.2}.

\subsection{Code mapping construction}\label{section4.1}

\par According to the proposed code mapping strategy, it is possible for each NFRS to be mapped to multiple VLCs. For ease of description, we introduced a $1\times N$ sequence called the \emph{mapping sequence} as $\bm{x}=(x_1,x_2,\dots,x_{N})$, where $N$ is the number of NFRSs in the original bitstream, and $x_{i}$ represents the number of VLCs mapped to the $i$-th NFRS. The mapping sequence $\bm{x}$ can be used to represent the constructed code mapping. Therefore, constructing an optimal code mapping is equivalent to finding the optimal mapping sequence, which can be considered a combinatorial optimization problem with constraints. The optimization objective is to minimize the file size increment with respect to the required capacity. A key constraint is that the capacity calculated by the mapping sequence $\bm{x}$ must be greater than or equal to the required capacity. In addition, according to the Huffman coding rules of the JPEG standard \cite{standard1992information}, two other constraints must be considered.
\begin{enumerate}
    \item For each NFRS, at least one VLC must be mapped to provide error-free encoding and decoding. Therefore, each element in the mapping sequence $\bm{x}$ must be an integer greater than or equal to one.
    \item According to Section K.2 in the JPEG standard \cite{standard1992information}, the number of customized VLCs, i.e., the sum of mapping sequence $\bm{x}$, can be up to 256. Therefore, the sum of $\bm{x}$ has a range of $N$ to 256.
\end{enumerate}
\par Let $z$ denote the required capacity, the optimization problem can be formulated as
\begin{equation}
    \begin{split}
        & \mathrm{minimize}\ \ \ \mathcal{F}(\bm{x}) \\
        & \mathrm{subject\ to}\
        \begin{cases}
            \mathcal{C}(\boldsymbol{x}) \geq z   & \\
            x_i\geq1, i \in [1,N] & \\
            \sum_{i=1}^N x_i\leq 256
        \end{cases}
    \end{split},
    \label{eq_problem}
\end{equation}
\noindent where $\mathcal{F}(\bm{x})$ represents the file size of the part of  marked bitstream that will change with different mapping sequences. Thus, minimizing the file size increment is equivalent to determining a mapping sequence with a minimal value of $\mathcal{F}(\bm{x})$. $\mathcal{C}(\boldsymbol{x})$ represents the embedding capacity achieved by the mapping sequence $\bm{x}$, which can be calculated as
\begin{equation}
    \mathcal{C}(\boldsymbol{x})=\sum_{i=1}^{N}f_i\cdot \left \lfloor \log_2 x_i  \right \rfloor,
    \label{eq_cal_capacity}
\end{equation}
\noindent where $f_i$ denotes the frequency of $i$-th NFRS in the original bitstream, and $\left\lfloor\cdot\right\rfloor$ is the floor operator.

\par In the workflow of the proposed framework, only the DHT segment and the entropy-coded data are changed; therefore, we only need to compare the file size change in the two parts. We note that when calculating the file size increment, the file size changed owing to byte alignment, and zero-byte padding was not included because the modification could not be calculated before ending the encoding, and it was extremely small that it could be negligible. 

\subsubsection{File size change in the DHT segment}

\par
\begin{figure*}[!ht]
    \centering
    \includegraphics[width = 1\textwidth]{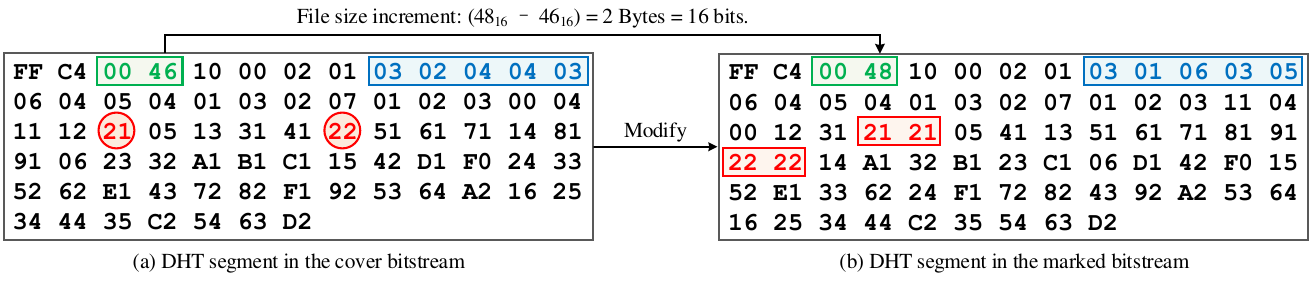}
    \caption{Comparison of the DHT segment before and after modification. The segment length is changed from \texttt{0x0046} to \texttt{0x0048}, which are marked in green. In the BITS list, five bytes are modified, which are marked in blue. In the HUFFVAL list, two RSs, ``2/1'' and ``2/2'', are assigned with two VLCs to construct the code mapping, which are marked in red.}
    \label{fig_dht_modification}
\end{figure*}

 The file size change in the DHT segment occurs in the DHT segment modification step, which is caused by adding the copies of the NFRSs to the HUFFVAL list to represent the code mapping. The original lists, BITS and HUFFVAL, were replaced with the lists generated during the DHT segment modification. Fig. \ref{fig_dht_modification} presents an example of the DHT segment before and after modification. As shown in Fig. \ref{fig_dht_modification}, the segment length is changed from \texttt{0x0046} to \texttt{0x0048}, which means the file size increment in the DHT segment is 16 bits. In this example, the RSs ``2/1'' and ``2/2'' are added to one copy to construct the code mapping. Therefore, in the HUFFVAL list, two extra RSs are added, which are marked in red. The BITS list was also modified to generate a customized Huffman table. We note that the values of the other RSs in the HUFFVAL list remained unchanged; however, the positions were adjusted to improve the compression ratio.
\par As the file sizes of other parts in the DHT segment are unchanged, calculating the file size change in the DHT segment is equivalent to calculating the file size of the new HUFFVAL list (in bits). Let $N^{\prime}$ denote the number of customized VLCs, which is calculated as
\begin{equation}
    N^{\prime} = \sum_{i=1}^{N}x_i.
\end{equation}
\noindent Because an RS occupies one byte (8 bits) of space in the HUFFVAL list, the file size change in the DHT segment $\mathcal{F}_{\mathrm{huffval}}$ can be calculated by 
\begin{equation}
    \mathcal{F}_{\mathrm{huffval}} = 8N^{\prime}.
    \label{eq_fs_huf}
\end{equation}

\subsubsection{File size change in the entropy-coded Data}

\par The file size change in the entropy-coded data is caused by VLC replacement. During the VLC replacement, all the original VLCs are substituted by the VLCs defined by the customized Huffman table. Thus, calculating the file size change in entropy-coded data is equivalent to calculating the file size of VLCs after replacement based on the customized Huffman table. However, customizing the Huffman table requires determining the optimal mapping sequence and, thus, the customized Huffman table cannot be obtained during the code mapping construction step. To this end, we propose a surrogate method from the perspective of information theory, which can approximate the file size of VLCs without a customized Huffman table.
\par According to information theory, the length of a VLC is close to its self-information (in bits) as the VLC is a Huffman code. Therefore, we can obtain the approximate length of the VLC by calculating its self-information. Calculating self-information requires the actual post-embedding frequency of VLCs. Considering that the actual post-embedding frequency cannot be obtained before data embedding, the post-embedding frequency can be estimated as follows:
\begin{equation}
    f_i^{\prime}\approx\left\{\begin{array}{ll}
        f_1/x_1,   & \text { if } 1 \leqslant i\leqslant x_1                         \\
        f_2/x_2,   & \text { if } x_1 < i\leqslant x_1+x_2                           \\
        \cdots                                                                              \\
        f_{N}/x_N, & \text { if } \sum_{j=1}^{N-1}x_j < i\leqslant N^{\prime}
    \end{array}\right.,
\end{equation}  
\noindent where $f_i^{\prime}$ represents the estimated post-embedding frequency of the $i$-th VLC in the customized Huffman table. This estimation is based on the fact that additional data are encrypted before embedding. That means the ``0'' and ``1'' in the additional data are distributed equally; therefore, the frequency of each VLC mapped to the same NFRS will be approximately identical.
\par The self-information of the $i$-th VLC $I_i$ can therefore be calculated as
\begin{equation}
    I_i=\log_2 \left (\frac{\sum_{j=1}^{N^{\prime}}f_j^{\prime}}{f_i^{\prime}}\right ).
    \label{eq_self_information}
\end{equation}
\par Finally, the file size of the VLCs, namely, $\mathcal{F}_{\mathrm{vlc}}$, can be estimated as
\begin{equation}
    \begin{split}
        \mathcal{F}_{\mathrm{vlc}} & =\sum_{i=1}^{N^{\prime}} I_i \cdot f_i^{\prime} = \sum_{i=1}^{N^{\prime}} \log_2\left (\frac{\sum_{j=1}^{N^{\prime}}f_j^{\prime}}{f_i^{\prime}}\right )\cdot f_i^{\prime}.
    \end{split}
    \label{eq_fs_vlc}
\end{equation}
\par In Eq. (\ref{eq_fs_vlc}), we multiply the estimated post-embedding frequency of VLC with its corresponding self-information and sum it to approximate the actual file size of VLCs. 
\par According to Eqs. (\ref{eq_fs_huf}) and (\ref{eq_fs_vlc}), the function $\mathcal{F}(\bm{x})$ can be written as:
\begin{equation}
    \begin{split}
        \mathcal{F}(\bm{x}) & = \mathcal{F}_{\mathrm{huffval}} + \mathcal{F}_{\mathrm{vlc}}. \\
        % & = 8N^{\prime}+ \sum_{i=1}^{N^{\prime}} \log_2\left (\frac{\sum_{j=1}^{N^{\prime}}f_j^{\prime}}{f_i^{\prime}}\right )\cdot f_i^{\prime}.
    \end{split}
    \label{eq_estimated_fs}
\end{equation}

\subsection{Huffman Table Customization}\label{section4.2}

\begin{figure*}[!ht]
    \centering
    \includegraphics[width = 1\textwidth]{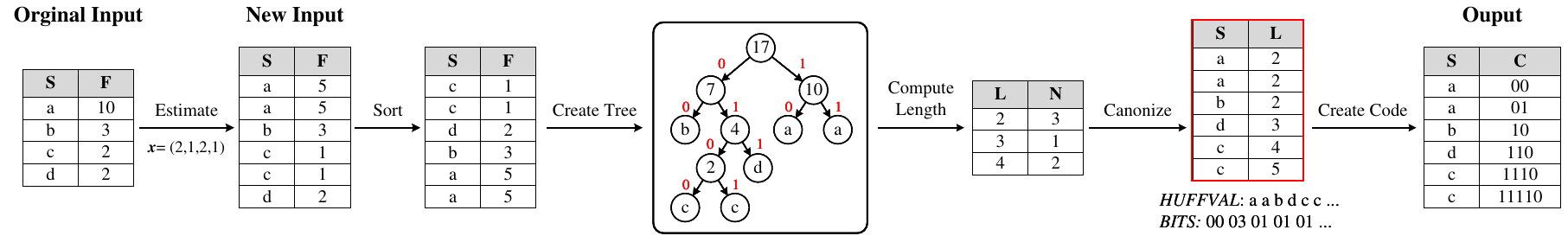}
    \caption{Example to generate the customized Huffman table. \textbf{S}: RS. \textbf{F}: frequency. \textbf{L}: length of VLC. \textbf{N}: number of VLC with the same length. \textbf{C}: VLC. }
    \label{fig_che}
\end{figure*}

\par Because the number of customized VLCs is flexible, the Huffman table customization step is different from previous CM-based methods; however, it is similar to the generation of the optimized Huffman table. The difference is that we considered the estimated post-embedding frequency as the input instead of the original RS frequency. Fig. \ref{fig_che} depicts a simplified example to illustrate the generation of the customized Huffman table. We note that a detailed version of the generation of the Huffman table can be found in Section K.2 of the JPEG standard \cite{standard1992information}. First, the original RS and corresponding frequency are transformed into a 2-tuple sequence $\{(a,10),(b,3),(c,2),(d,2)\}$ to generate a new input. Assuming that the mapping sequence is $\bm{x}=(2,1,2,1)$, the new 2-tuple sequence is then generated by estimating the post-embedding frequency, that is, $\{(a,5),(a,5),(b,3),(c,1),(c,1),(d,2)\}$. After sorting the RSs according to frequency, the Huffman tree was created and used to compute the code lengths. The codes were then canonized to create a canonical Huffman code. Canonical Huffman code lengths were recorded on the BITS list. As shown in Fig. \ref{fig_che}, the numbers of code lengths 2, 3, 4, and 5 are 3, 1, 1, and 1, respectively. Therefore, the BITS list is ``00 03 01 01 01...". Next, a high-frequency RS is assigned with a short code to compress the bitstream. The HUFFVAL list was created according to the code length. These two lists were used to generate a customized Huffman table.

\section{Example: a new CM-based RDH method}\label{section5}
% with some problem-oriented designs to improve search accuracy and convergence speed.
\par This section presents a new CM-based RDH method to realize the proposed framework. First, a GA is introduced to solve the optimization problem; thus, a nearly optimal code mapping is constructed. Then, the data embedding and extraction procedures are described.

\subsection{Problem solving using GA}

\begin{algorithm}[!ht]
    \caption{GA procedure}
    \LinesNumbered
    \label{alg_GA}
    \KwIn{\\
        \quad Population size $K$, iterations $I$, crossover rate $R_c$, \\
        \quad and mutation rate $R_m$}
    \KwOut{\\
        \quad The best individual $p^*$
    }
    Generate the initial parent population of $K$ individuals, $P=\{p_1,p_2,\dots,p_K\}$\;
    \For{$i=1$ to $I$}{
        Compute the fitness of each $p_i\in P$\;
        Find the best individual and define it as $p^*$\;
        Insert $p^*$ into the new population $P^{\prime}$.
        Select $K-1$ individuals from $P$ for reproduction. \;
        The crossover operation is performed based on $R_c$.
        Perform a mutation operation based on $R_m$\;
        Insert the new $K-1$ individuals into $P^{\prime}$\;
        $P\leftarrow P^{\prime}$\;
    }
    \Return $p^*$
\end{algorithm}
Evaluating all mapping sequences is impractical, because the solution space is very large. Thus, designing a fast and effective optimization algorithm is vital for searching for a nearly optimal solution. We leveraged the GA to automatically search for a nearly optimal solution. We designed several specific problem-oriented rules to improve search accuracy and convergence speed. Algorithm \ref{alg_GA} describes the GA procedure. In this algorithm, $K$ individuals are randomly generated to initialize the parent population $P=\{p_1,p_2,\dots,p_K\}$. Subsequently, the fitness of each individual in $P$ was computed. To complete the reproduction of the new population, GA operations (i.e., selection, crossover, and mutation) were performed to generate new individuals. The reproduction of the new population terminates when the maximum number of iterations $I$ is reached.

\subsubsection{Population initialization}

\begin{figure}[!ht]
    \centering
    \includegraphics[width= 0.4\textwidth]{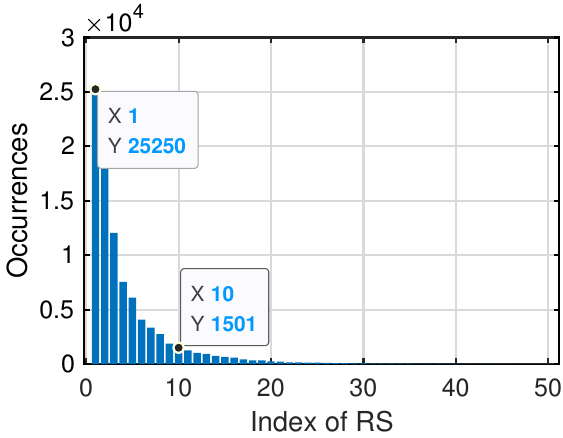}
    \caption{Distribution of RSs of the Baboon image under a QF of 70.}
    \label{freq_baboon}
\end{figure}

\par One mapping sequence $\bm{x}$ is regarded as an individual of a population. This means that $N$ genes are encoded in an individual when the number of NFRSs is $N$. Each gene corresponds to an element in mapping sequence $\bm{x}$. However, not all NFRSs are suitable for assigning multiple VLCs. Some low-frequency RSs contribute only to the limited embedding capacity when mapping multiple VLCs to them. Fig. \ref{freq_baboon} illustrates the RS frequency of the image Baboon with a quality factor (QF) of 70. As shown in Fig. \ref{freq_baboon}, the frequencies of most RSs are low and not suitable for constructing code mapping when embedding a large payload of additional data. When searching for a nearly optimal solution, these RSs will occupy a portion of the available VLCs and influence the convergence speed. With this concern in mind, we selected a certain number of RSs to construct code mapping rather than all RSs. After several preliminary trials (not reported here), we finally considered ten as the number in the proposed method. We limited the selected 10 RSs to consecutive ones, where the leftmost one was closest to and greater than the required capacity. The RSs may encounter the following situations:1) less than 10 RSs are available and 2) the leftmost RS is smaller than the required capacity. When the number of RSs is less than 10, all RSs are selected. When the leftmost capacity is less than the required capacity, the RS is selected starting from the leftmost one.
l
\subsubsection{Individual encoding}

\par To solve the optimization problem using the GA, all the tunable parameters in an individual, i.e., the mapping sequence $\bm{x}=(x_1,x_2,\dots,x_N)$ should be represented with binary encoding. Considering that using three or more bits to represent an element in $\bm{x}$ will decelerate the convergence and produce more infeasible solutions, we limit the value of each $x_i$ to one in $\{1, 2, 4, 8\}$ to reduce the solution space. Each value in $\{1,2,4,8\}$ occupies two bits, such as the codes ``00'', ``01'', ``10'', and ``11'', which are used to represent 1, 2, 4, 8, respectively. Based on this rule, one VLC can represent up to three bits of data ($\log_2 8$). Thus, an individual is composed of a sequence of 20 bits. Therefore, the $i$th individual, $p_i$ in the population can be represented by
\begin{equation}
    {p_i} = \{a_1^1a_1^2\cdots a_{10}^1a_{10}^2 \mid a_j^1,a_j^2 \in \{0,1\} \ \text{and}\ j  \in [1,10]\},
\end{equation}
\noindent where $(a_j^1,a_j^2)$ represents the $j$-th genes in an individual.

\subsubsection{Selection, crossover, and mutation}

\par In the reproduction process of the population, we adopted the elite preservation strategy. We inserted it into the new population for the best individual instead of modifying it. We used roulette wheel selection to select better individuals for the remaining individuals. After the crossover and mutation steps, the newly generated individuals are inserted into the next generation.

\subsubsection{Fitness function}

\par The fitness function is used to evaluate the quality of individuals. This eventually causes GA to evolve toward an individual with better fitness. According to the optimization problem formulated in Eq. (\ref{eq_problem}), the optimal individual has the minimum value of $\mathcal{F}(\bm{x})$. However, roulette-wheel selection selects individuals with higher fitness values. Thus, we adjust the fitness value for the $i$th individual as follows:
\begin{equation}
    fitness(i) = \mathcal{F}_\mathrm{max}-\mathcal{F}_i,
\end{equation}
\noindent where $\mathcal{F}_\mathrm{max}$ denotes the maximal value of $\mathcal{F}(\bm{x})$ in one generation, and $\mathcal{F}_i$ is the value of $\mathcal{F}(\bm{x})$ calculated by the $i$-th individual. The fitness value is set to zero as a penalty for an individual who does not satisfy the capacity constraint.

\subsection{Data embedding \& extraction}

\begin{figure}[!ht]
    \centering
    \includegraphics[width= 1\textwidth]{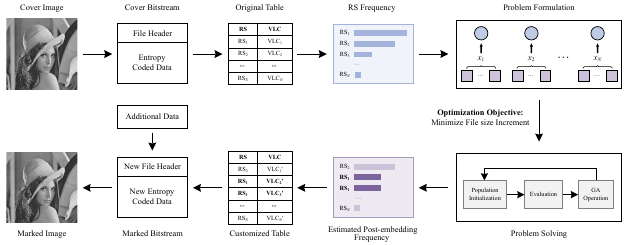}
    \caption{Embedding process of the proposed method.}
    \label{fig_framework}
\end{figure}

\par Fig. \ref{fig_framework} depicts the embedding procedures of the proposed method, which are as follows: 
\begin{enumerate}
    \item Construct the Huffman table by parsing the file header in the bitstream.
    \item Count the RS frequency by parsing the entropy-coded data.
    \item Convert the code mapping construction to a combinatorial optimization problem and solve it using the GA.
    \item Calculate the estimated post-embedding frequency after obtaining the nearly optimal solution, and construct the customized Huffman table
    \item Embed the additional data into the entropy-coded data and modify the file header.
    \item Generate the marked JPEG bitstream by combining the new entropy-coded data with the new file header. 
\end{enumerate}

\par
\begin{figure*}[htbp]
    \centering
    \includegraphics[width = 1\textwidth]{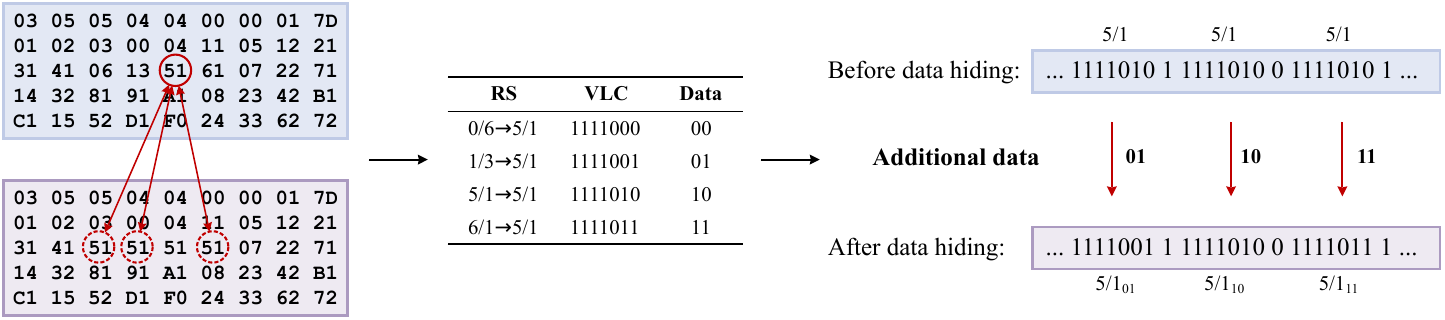}
    \caption{CM-based data embedding example. }
    \label{fig_embedding_example}
\end{figure*}

 Fig. \ref{fig_embedding_example} depicts a basic data embedding example by code mapping. We assume that ``0/6'', ``1/3'', and ``6/1'' are all zero-frequency RSs. The corresponding VLCs are unused and can be used to construct mapping relationships. After modifying the three RSs to ``5/1'', the code mapping is constructed, and each VLC represents two bits of different data. The additional data ``011011'' can then be embedded by replacing the original VLCs with VLCs in the mapping set.

\par The code mapping constructed in the data embedding process is required to extract the additional data embedded in the marked JPEG bitstream. Instead of solving the optimization problem, code mapping can be directly constructed by parsing the DHT segment during data extraction. The entropy-coded data are further parsed and additional data can then be extracted. Because the marked image is lossless to the original image in terms of the image quality, the image content does not need to be restored. The customized Huffman table must be replaced by different Huffman tables to restore the bitstream. The standard Huffman table can be found in the JPEG standard \cite{standard1992information}, and an optimized Huffman table can be generated according to the RS frequency. The main data extraction procedure is described as follows: 
\begin{enumerate}
    \item Generate the customized Huffman table for AC coefficients by parsing the DHT segment from the marked JPEG bitstream.
    \item Construct the mapping relationship according to the customized Huffman table.
    \item Extract the embedded data from the entropy-coded data according to the mapping relationship.
    \item Synthesize a new JPEG bitstream using the optimized Huffman table or standard Huffman table. 
\end{enumerate}

\section{Experimental results}\label{section6}

\par In this section, we first introduce the experimental settings in Section \ref{section6.1}. The results and comparisons with previous methods in terms of file size preservation, image quality, and running time are presented in Sections \ref{section6.2}, \ref{section6.3}, and \ref{section6.4}, respectively.

\subsection{Experimental settings}\label{section6.1}

\subsubsection{Datasets}

\par The images tested in our experiments are selected from two commonly used image databases:

\begin{itemize}
    \item \textbf{USC-SIPI} \cite{usc-sipi} is a collection of digitized images. We select four classical $512 \times 512$ images, Lena, Elaine, Baboon, and Boat from this image database. These selected images after JPEG compression have a QF = 70. The four selected images can be classified into two categories. The first two images, Lena and Elaine, contain more low-frequency details and smooth areas. More high-frequency details and texture areas are contained in the other two images, Baboon and Boat.
    \item \textbf{BOSSbase} \cite{bossbase} consists of 10000 512 $\times$ 512 grayscale images in the PGM format. For follow-up experiments, 200 images from the BOSSbase image database were randomly selected.
\end{itemize}

\par Because the images in the two databases are uncompressed images, we converted all the selected images into grayscale JPEG images. To convert the image from the uncompressed format, TIFF or PGM, into the JPEG format, the MATLAB \texttt{imwrite} function was used. All JPEG images generated by the \texttt{imwrite} function were encoded using the standard Huffman table. To verify the applicability of our proposed method, the JPEG images encoded with the optimized Huffman table were regarded as the original images in our experiments. JPEG images encoded with the optimized Huffman table were generated using the \texttt{jpeg\_read} and \texttt{jpeg\_write} functions in the JPEG toolbox \cite{jpeg_toolbox}.

\subsubsection{Evaluation metrics}

\par To evaluate the performance of the proposed method, three criteria, including the file-size increment, the mean square error (MSE) and the running time, are employed:
\begin{itemize}
    \item \textbf{File-size Increment (FI)} is the difference of file-sizes of the marked bitstream and the original bitstream, which is used to evaluate the file size preservation performance.
    \item  \textbf{Mean Square Error (MSE)} is employed to evaluate the image quality of the marked JPEG image, which is calculated as
    \begin{equation}
        MSE=\frac{1}{h\times w}  { \sum_{i=1}^{h}} {\sum_{j=1}^{w}} \left ( X_{i,j}-X^{\prime}_{i,j} \right )^2, 
        \label{eq_mse}
    \end{equation}
    \noindent where $h$ and $w$ are the height and width of original JPEG image, respectively. $X_{i,j}$ and $X^{\prime}_{i,j}$ are the $(i,j)$-th pixel in original JPEG image and marked JPEG image, respectively. According to Eq. (\ref{eq_mse}), the smaller MSE represents less distortion and better visual quality. When the MSE is equal to zero, the marked JPEG image is lossless compared to the original JPEG image.
    \item \textbf{Running Time} refers to the running time of the data embedding process. We use it to evaluate the complexity of our chosen methods. For a fair comparison, we tested all the methods in the same environment, including the software and hardware. 
\end{itemize}

\subsubsection{Baselines}
\par To validate the effectiveness of our proposed method, three state-of-the-art high capacity CM-based RDH methods were selected for comparison, including method \cite{qiu2020optimized} (TCSVT2021), method \cite{du2020gvm} (TDSC2020), and method \cite{zhang2020reversible} (JVCIR2020). In \cite{qiu2020optimized}, Qiu \emph{et~al.} first proposed a basic RDH method to preserve file size. The embedding capacity achieved by the basic method outperforms previous file size-preserving methods \cite{qian2012lossless,hu2013improved,qiu2018lossless}. They also proposed an extension method to achieve higher capacity while increasing the file size. In subsequent experiments, both methods were chosen for comparison. In addition, some state-of-the-art HS-based RDH methods were selected to compare the performance of file size preservation, visual quality, and time complexity with our proposed method, including \cite{he2020reversible} (TIFS2020), \cite{li2020reversible} (SP2020), \cite{yin2020reversible} (TCSVT2020), and \cite{xiao2020efficient} (TCSVT2021).

\subsubsection{Implementation details}
\par We performed the experiments for all the state-of-the-art RDH methods \cite{du2020gvm,qiu2020optimized,zhang2020reversible,he2020reversible,li2020reversible,yin2020reversible,xiao2020efficient} and the proposed method on a PC running Windows 10 with 16 GB RAM and an 8-core AMD Ryzen 7-2700 processor. Notably, the original JPEG images must be encoded using the standard Huffman table in the methods \cite{du2020gvm,qiu2020optimized,zhang2020reversible,yin2020reversible}. In the method proposed by He \emph{et~al.} \cite{he2020reversible}, the weighting factor $\alpha$ of the negative influence model was set to zero and one. $\alpha=0$ indicates that only visual distortion was considered. Hence, the visual quality performance was better than that of the other values. $\alpha=1$ indicates that the file size preservation performance is better than that of the other values. 
\par In this paper, the parameters of the GA in the proposed method are determined based on preliminary experiments, e.g., population size $K$, max iterations $I$, crossover rate $R_c$, and mutation rate $R_m$, are set to 100, 50, 0.8, and 0.3, respectively.

\subsection{Evaluation on file size preservation}\label{section6.2}

\par To compare the performance of file-size preservation with CM-based methods, we first obtained the embedding capacity (EC) achieved by Qiu \emph{et~al.} \cite{qiu2020optimized} with different images and different QFs. We tested the results of the two methods proposed in \cite{qiu2020optimized}, that is, the basic RDH method and the extension method. When calculating the embedding capacity for the extension method, we performed one VLC relay transfer before performing the basic method, as described in Section III-B in \cite{qiu2020optimized}. Subsequently, we embedded the same quantity of additional data into the test images by using different CM-based methods. We note that the actual EC obtained by the method \cite{zhang2020reversible} depends on a threshold $T$ and the required capacity. In our experiments, the threshold $T$ was set to six, and the required capacity was set to the EC obtained by the extension method in \cite{qiu2020optimized}. The comparison results of FIs with previous CM-based RDH methods \cite{qiu2020optimized,du2020gvm,zhang2020reversible} are listed in Table. \ref{tab:fi_cm}, in which TCSVT2021-I represents the basic RDH method and TCSVT2021-II represents the extension method. The original JPEG images were encoded using a standard Huffman table. As shown in the Table. \ref{tab:fi_cm}, the FIs of our method are the smallest in all cases, which means our proposed method preserves the file size better than previous state-of-the-art CM-based methods. Some results of our method are even negative, that is, the file size of the marked bitstream is smaller than that of the original JPEG bitstream.

\begin{table}[H]
    \centering
    \caption{FI (bits) of test images with different QFs using the proposed method and previous CM-based methods \cite{du2020gvm, qiu2020optimized,zhang2020reversible}. The original JPEG images are encoded with the standard Huffman table.}
    \label{tab:fi_cm}
    \resizebox{\textwidth}{!}{%
    \begin{tabular}{@{}llrrrrrrrrrrrrrrrrrr@{}}
    \toprule
    \multirow{2}{*}{\textbf{Image}} & \multirow{2}{*}{\textbf{Method}} & \multicolumn{2}{c}{\textbf{QF=10}} & \multicolumn{2}{c}{\textbf{QF=20}} & \multicolumn{2}{c}{\textbf{QF=30}} & \multicolumn{2}{c}{\textbf{QF=40}} & \multicolumn{2}{c}{\textbf{QF=50}} & \multicolumn{2}{c}{\textbf{QF=60}} & \multicolumn{2}{c}{\textbf{QF=70}} & \multicolumn{2}{c}{\textbf{QF=80}} & \multicolumn{2}{c}{\textbf{QF=90}} \\ \cmidrule(l){3-20} 
     &  & \multicolumn{1}{c}{EC} & \multicolumn{1}{c}{FI} & \multicolumn{1}{c}{EC} & \multicolumn{1}{c}{FI} & \multicolumn{1}{c}{EC} & \multicolumn{1}{c}{FI} & \multicolumn{1}{c}{EC} & \multicolumn{1}{c}{FI} & \multicolumn{1}{c}{EC} & \multicolumn{1}{c}{FI} & \multicolumn{1}{c}{EC} & \multicolumn{1}{c}{FI} & \multicolumn{1}{c}{EC} & \multicolumn{1}{c}{FI} & \multicolumn{1}{c}{EC} & \multicolumn{1}{c}{FI} & \multicolumn{1}{c}{EC} & \multicolumn{1}{c}{FI} \\ \midrule
    Baboon & TCSVT2021-I \cite{qiu2020optimized} & 7222 & 168 & 3974 & 256 & 2237 & 256 & 2262 & 416 & 1740 & 320 & 1786 & 496 & 1831 & 888 & 1219 & 280 & 1124 & 648 \\
     & Proposed & 7222 & \textbf{-7296} & 3974 & \textbf{-7480} & 2237 & \textbf{-6864} & 2262 & \textbf{-5144} & 1740 & \textbf{-4424} & 1786 & \textbf{-3704} & 1831 & \textbf{-2080} & 1219 & \textbf{-2640} & 1124 & \textbf{-8880} \\ \cmidrule(l){2-20} 
     & TCSVT2021-II \cite{qiu2020optimized} & 10691 & 2168 & 15166 & 8360 & 17925 & 14544 & 20239 & 19424 & 21761 & 24792 & 23892 & 30896 & 26419 & 37712 & 30913 & 46264 & 39899 & 60112 \\
     & TDSC2020  \cite{du2020gvm} & 10691 & -392 & 15166 & 11280 & 17925 & 16824 & 20239 & 21688 & 21761 & 24552 & 23892 & 31224 & 26419 & 39152 & 30913 & 43272 & 39899 & 45928 \\
     & Proposed & 10691 & \textbf{-3736} & 15166 & \textbf{3720} & 17925 & \textbf{8952} & 20239 & \textbf{12904} & 21761 & \textbf{15632} & 23892 & \textbf{19032} & 26419 & \textbf{22712} & 30913 & \textbf{27624} & 39899 & \textbf{30232} \\ \cmidrule(l){2-20} 
     & JVCIR2020 \cite{zhang2020reversible} & 8930 & 15887 & 12105 & 32219 & 13813 & 37813 & 10145 & 14328 & 15190 & 46369 & 17619 & 52316 & 12805 & 18407 & 22937 & 66247 & 21520 & 70103 \\
     & Proposed & 8930 & \textbf{-5584} & 12105 & \textbf{656} & 13813 & \textbf{4760} & 10145 & \textbf{2888} & 15190 & \textbf{8960} & 17619 & \textbf{12296} & 12805 & \textbf{9024} & 22937 & \textbf{19488} & 21520 & \textbf{11440} \\ \midrule
    Boat & TCSVT2021-I \cite{qiu2020optimized} & 1656 & 16 & 1219 & -16 & 1307 & 16 & 1139 & 88 & 1228 & 128 & 1332 & 160 & 1154 & 353 & 1424 & 200 & 2338 & 992 \\
     & Proposed & 1657 & \textbf{-8672} & 1219 & \textbf{-6760} & 1307 & \textbf{-4416} & 1139 & \textbf{-3248} & 1228 & \textbf{-1648} & 1332 & \textbf{-1192} & 1154 & \textbf{-1872} & 1424 & \textbf{-4200} & 2338 & \textbf{-12544} \\ \cmidrule(l){2-20} 
     & TCSVT2021-II \cite{qiu2020optimized} & 4755 & 1584 & 7697 & 4568 & 9984 & 6976 & 11354 & 10272 & 12613 & 13352 & 13939 & 16368 & 15831 & 20696 & 19459 & 27744 & 27887 & 38024 \\
     & TDSC2020  \cite{du2020gvm} & 4755 & -3696 & 7697 & 2680 & 9984 & 7840 & 11354 & 13080 & 12613 & 15728 & 13939 & 19456 & 15831 & 24200 & 19459 & 29256 & 27887 & 40304 \\
     & Proposed & 4755 & \textbf{-5512} & 7697 & \textbf{-360} & 9984 & \textbf{4280} & 11354 & \textbf{7000} & 12613 & \textbf{9328} & 13939 & \textbf{11448} & 15831 & \textbf{13048} & 19459 & \textbf{14160} & 27887 & \textbf{12928} \\ \cmidrule(l){2-20} 
     & JVCIR2020 \cite{zhang2020reversible} & 3882 & 9653 & 5647 & 16189 & 5062 & 15405 & 6107 & 18949 & 5793 & 17702 & 13641 & 27892 & 11791 & 36597 & 14883 & 45350 & 21436 & 67137 \\
     & Proposed & 3882 & \textbf{-6424} & 5647 & \textbf{-2336} & 5062 & \textbf{-528} & 6107 & \textbf{1704} & 5793 & \textbf{2800} & 13641 & \textbf{10992} & 11791 & \textbf{9064} & 14883 & \textbf{9360} & 21436 & \textbf{6624} \\ \midrule
    Elaine & TCSVT2021-I \cite{qiu2020optimized} & 1056 & -8 & 966 & 8 & 877 & -16 & 1021 & -8 & 1114 & -144 & 895 & 88 & 951 & 296 & 1196 & 184 & 2034 & 552 \\
     & Proposed & 1056 & \textbf{-9928} & 966 & \textbf{-7528} & 877 & \textbf{-6240} & 1021 & \textbf{-5496} & 1114 & \textbf{-6048} & 895 & \textbf{-6688} & 951 & \textbf{-6288} & 1196 & \textbf{-8880} & 2034 & \textbf{-12480} \\ \cmidrule(l){2-20} 
     & TCSVT2021-II \cite{qiu2020optimized} & 3384 & 1008 & 5316 & 2832 & 6672 & 4624 & 8103 & 6328 & 9614 & 8064 & 11595 & 10008 & 14843 & 13672 & 20436 & 19992 & 33306 & 37528 \\
     & TDSC2020  \cite{du2020gvm} & 3384 & -5008 & 5316 & -1520 & 6672 & 2048 & 8103 & 5672 & 9614 & 7800 & 11595 & 11080 & 14843 & 15936 & 20436 & 29376 & 33306 & 52656 \\
     & Proposed & 3384 & \textbf{-7592} & 5316 & \textbf{-3096} & 6672 & \textbf{-232} & 8103 & \textbf{1568} & 9614 & \textbf{2512} & 11595 & \textbf{4024} & 14843 & \textbf{7480} & 20436 & \textbf{10440} & 33306 & \textbf{19120} \\ \cmidrule(l){2-20} 
     & JVCIR2020 \cite{zhang2020reversible} & 1507 & 3547 & 3292 & 8516 & 3634 & 12069 & 7164 & 20012 & 9181 & 26631 & 11561 & 34727 & 10666 & 34852 & 15670 & 48734 & 31865 & 95755 \\
     & Proposed & 1507 & \textbf{-9464} & 3292 & \textbf{-5088} & 3634 & \textbf{-3480} & 7164 & \textbf{664} & 9181 & \textbf{2072} & 11561 & 3912 & \textbf{10666} & 3528 & 15670 & \textbf{5520} & 31865 & \textbf{17592} \\ \midrule
    Lena & TCSVT2021-I \cite{qiu2020optimized} & 1275 & 8 & 1032 & -8 & 948 & -72 & 972 & 0 & 1033 & 40 & 827 & 16 & 867 & 32 & 878 & 136 & 1051 & 302 \\
     & Proposed & 1275 & \textbf{-8440} & 1032 & \textbf{-6704} & 948 & \textbf{-5072} & 972 & \textbf{-3576} & 1033 & \textbf{-2752} & 827 & \textbf{-2568} & 867 & \textbf{-2312} & 878 & \textbf{-1856} & 1051 & \textbf{-3696} \\ \cmidrule(l){2-20} 
     & TCSVT2021-II \cite{qiu2020optimized} & 3607 & 1480 & 5424 & 3504 & 6870 & 5584 & 7978 & 7464 & 9084 & 9272 & 10279 & 11280 & 12173 & 14520 & 15468 & 19384 & 22988 & 30688 \\
     & TDSC2020  \cite{du2020gvm} & 3607 & -4496 & 5424 & -40 & 6870 & 3888 & 7978 & 6184 & 9084 & 10928 & 10279 & 12272 & 12173 & 15640 & 15468 & 21992 & 22988 & 39256 \\
     & Proposed & 3607 & \textbf{-6024} & 5424 & \textbf{-2224} & 6870 & \textbf{920} & 7978 & \textbf{3448} & 9084 & \textbf{5272} & 10279 & \textbf{7040} & 12173 & \textbf{9152} & 15468 & \textbf{12752} & 22988 & \textbf{18168} \\ \cmidrule(l){2-20} 
     & JVCIR2020 \cite{zhang2020reversible} & 3455 & 6091 & 4289 & 12094 & 4936 & 14320 & 5415 & 17373 & 6665 & 19893 & 8457 & 25288 & 8826 & 28454 & 9813 & 33808 & 14366 & 46473 \\
     & Proposed & 3455 & \textbf{-6256} & 4289 & \textbf{-3352} & 4936 & \textbf{-1064} & 5415 & \textbf{832} & 6665 & \textbf{3056} & 8457 & \textbf{5168} & 8826 & \textbf{5632} & 9813 & \textbf{7104} & 14366 & \textbf{9808} \\ \bottomrule
    \end{tabular}%
    }
\end{table}

\par We also compared the FIs with the state-of-the-art HS-based RDH methods. Four images from the USC-SIPI database were tested with different payloads and QFs. The FIs of the std-bitstream and opt-bitstream are different, and thus, they are all evaluated in this experiment. The comparison results for the std-bitstream are listed in Table. \ref{tab:fi_standard}. The comparison results for the opt-bitstream are listed in Table. \ref{tab:fi_optimized}. The smallest FI values are shown in bold in the two tables. The std-bitstream can be observed from the Table. \ref{tab:fi_standard}, which shows that the FIs obtained by our method are the smallest. For the opt-bitstream, it can be observed from Table. \ref{tab:fi_optimized} that most results obtained by the proposed method are smaller than those of the other HS-based methods. The results of the proposed method fluctuate near the given payload. This difference is related to the zero-byte padding and byte-alignment operations in the bitstream generation process.

\begin{table}[H]
    \centering
    \caption{FI (bits) of the marked JPEG images under different QFs and different payloads using the proposed method and the previous HS-based RDH methods. The original JPEG images are encoded with the standard Huffman table.}
    \label{tab:fi_standard}
    \resizebox{\textwidth}{!}{%
    \begin{tabular}{@{}llrrrrrrrrrrrrrrrr@{}}
    \toprule
    \multirow{2}{*}{\textbf{Image}} &
      \multirow{2}{*}{\textbf{Method}} &
      \multicolumn{4}{c}{\textbf{Payload (bits) with QF=30}} &
      \multicolumn{4}{c}{\textbf{Payload (bits) with QF=50}} &
      \multicolumn{4}{c}{\textbf{Payload (bits) with QF=70}} &
      \multicolumn{4}{c}{\textbf{Payload (bits) with QF=90}} \\ \cmidrule(l){3-18} 
           &                          & 2000 & 3000 & 4000 & 5000 & 3000 & 5000 & 7000  & 9000  & 4000 & 7000  & 10000 & 13000 & 5000 & 9000  & 13000 & 17000 \\ \midrule
    % Baboon & SP2018  \cite{hou2018reversible}           & 2896 & 3832 & 5168 & 6888 & 4016 & 6592 & 9184  & 11648 & 5616 & 10144 & 14424 & 19216 & 8160 & 14456 & 22160 & 29192 \\
    Baboon & TIFS2020 ($\alpha$=0) \cite{he2020reversible} & 2888 & 4040 & 5384 & 6520 & 3696 & 5856 & 8240  & 11840 & 5688 & 9632  & 13904 & 18672 & 7784 & 14272 & 21912 & 27920 \\
           & TIFS2020 ($\alpha$=1) \cite{he2020reversible} & 1536 & 2448 & 3688 & 4624 & 2616 & 4488 & 7000  & 9168  & 4008 & 6960  & 11152 & 15360 & 5016 & 9512  & 14224 & 19128 \\
           & SP2020 \cite{li2020reversible}             & 2112 & 2888 & 4192 & 5328 & 3192 & 5200 & 7472  & 9984  & 4936 & 8432  & 12384 & 16608 & 7200 & 12704 & 19120 & 24040 \\
           & TCSVT2020  \cite{yin2020reversible}           & 2296 & 3472 & 4664 & 5760 & 3568 & 5888 & 8576  & 11032 & 5216 & 9536  & 13880 & 18496 & 7856 & 13856 & 21072 & 27680 \\
           & TCSVT2021   \cite{xiao2020efficient}                & 2400 & 3528 & 5280 & 6008 & 3304 & 5720 & 8232  & 11160 & 4568 & 9016  & 12976 & 17224 & 6592 & 12384 & 19152 & 24536 \\
     &
      Proposed &
      \textbf{-7224} &
      \textbf{-6088} &
      \textbf{-5144} &
      \textbf{-4064} &
      \textbf{-3176} &
      \textbf{-1232} &
      \textbf{824} &
      \textbf{2656} &
      \textbf{32} &
      \textbf{3344} &
      \textbf{6368} &
      \textbf{9152} &
      \textbf{-5072} &
      \textbf{-912} &
      \textbf{3152} &
      \textbf{6896} \\ \midrule
    % Boat   & SP2018    \cite{hou2018reversible}         & 3120 & 4048 & 5832 & 6648 & 4104 & 7080 & 9800  & 12336 & 5720 & 10432 & 14360 & 18720 & 7360 & 14304 & 21480 & 28408 \\
    Boat       & TIFS2020 ($\alpha$=0) \cite{he2020reversible}& 2928 & 4184 & 5272 & 6408 & 4232 & 6544 & 9480  & 12232 & 5784 & 10272 & 14328 & 19264 & 6384 & 11952 & 18272 & 24760 \\
           & TIFS2020 ($\alpha$=1) \cite{he2020reversible}& 1896 & 3024 & 4144 & 5176 & 2920 & 5584 & 8248  & 11560 & 4888 & 8856  & 13248 & 17992 & 5408 & 10488 & 15784 & 21560 \\
           & SP2020    \cite{li2020reversible}          & 2296 & 3272 & 4360 & 5360 & 3472 & 5664 & 8032  & 10664 & 4344 & 8336  & 12320 & 16384 & 5024 & 10248 & 16064 & 21928 \\
           & TCSVT2020     \cite{yin2020reversible}        & 2464 & 3128 & 4920 & 5416 & 3296 & 5736 & 8264  & 11064 & 4744 & 9208  & 13536 & 17880 & 6928 & 12568 & 19184 & 25728 \\
           & TCSVT2021     \cite{xiao2020efficient}                & 2504 & 3608 & 5152 & 6144 & 4136 & 6408 & 8664  & 11744 & 5736 & 9216  & 13608 & 17840 & 5752 & 11112 & 16472 & 22888 \\
     &
      Proposed &
      \textbf{-3672} &
      \textbf{-2616} &
      \textbf{-1600} &
      \textbf{-704} &
      \textbf{-72} &
      \textbf{1864} &
      \textbf{3640} &
      \textbf{5800} &
      \textbf{1112} &
      \textbf{3920} &
      \textbf{7312} &
      \textbf{10104} &
      \textbf{-10088} &
      \textbf{-5784} &
      \textbf{-1784} &
      \textbf{2272} \\ \midrule
    % Elaine & SP2018    \cite{hou2018reversible}         & 3440 & 4800 & 6256 & 7568 & 4616 & 7600 & 10872 & 13624 & 6464 & 10816 & 14672 & 18960 & 7208 & 13704 & 20280 & 26304 \\
    Elaine       & TIFS2020 ($\alpha$=0) \cite{he2020reversible}& 3304 & 5168 & 6264 & 7400 & 4496 & 7456 & 9776  & 11992 & 5176 & 8992  & 12008 & 15512 & 6792 & 12072 & 17312 & 23160 \\
           & TIFS2020 ($\alpha$=1) \cite{he2020reversible}& 1752 & 2888 & 4056 & 5520 & 3008 & 5664 & 8120  & 10816 & 3480 & 7072  & 10464 & 14520 & 4144 & 7760  & 11976 & 16536 \\
           & SP2020      \cite{li2020reversible}        & 1960 & 2976 & 4096 & 5304 & 3616 & 5792 & 8120  & 10280 & 5184 & 8456  & 11416 & 14568 & 6392 & 11272 & 16784 & 22096 \\
           & TCSVT2020     \cite{yin2020reversible}        & 2640 & 4048 & 5360 & 6848 & 3776 & 6440 & 9632  & 12656 & 5160 & 9120  & 13440 & 17432 & 6872 & 12776 & 18632 & 24944 \\
           & TCSVT2021      \cite{xiao2020efficient}               & 3296 & 4456 & 6184 & 7256 & 4472 & 6256 & 9176  & 11368 & 4640 & 8160  & 11376 & 15224 & 6192 & 11000 & 16600 & 22608 \\
     &
      Proposed &
      \textbf{-5080} &
      \textbf{-4120} &
      \textbf{-2976} &
      \textbf{-1976} &
      \textbf{-4208} &
      \textbf{-2064} &
      \textbf{-120} &
      \textbf{1960} &
      \textbf{-3304} &
      \textbf{-152} &
      \textbf{2472} &
      \textbf{5656} &
      \textbf{-9344} &
      \textbf{-5312} &
      \textbf{-1216} &
      \textbf{2736} \\ \midrule
    % Lena   & SP2018   \cite{hou2018reversible}          & 2728 & 3784 & 5136 & 6664 & 4296 & 7008 & 9552  & 12424 & 5808 & 10080 & 14032 & 18176 & 6568 & 12208 & 17640 & 23648 \\
    Lena       & TIFS2020 ($\alpha$=0) \cite{he2020reversible}& 2824 & 4152 & 5184 & 6568 & 4392 & 6704 & 9440  & 12240 & 5384 & 9392  & 13184 & 17944 & 6344 & 10744 & 16064 & 21104 \\
           & TIFS2020 ($\alpha$=1) \cite{he2020reversible}& 1896 & 2864 & 4088 & 5528 & 3008 & 5368 & 8352  & 11240 & 3792 & 7856  & 12496 & 17352 & 4736 & 9840  & 14800 & 20432 \\
           & SP2020        \cite{li2020reversible}      & 2048 & 3400 & 4736 & 5928 & 3288 & 5616 & 8256  & 10896 & 4336 & 7736  & 11384 & 15688 & 5360 & 9408  & 14304 & 19072 \\
           & TCSVT2020       \cite{yin2020reversible}      & 2248 & 3016 & 4312 & 5624 & 3600 & 6344 & 8720  & 11640 & 5048 & 9232  & 12960 & 17744 & 6024 & 11336 & 16712 & 22848 \\
           & TCSVT2021   \cite{xiao2020efficient}                  & 2792 & 3744 & 4752 & 6408 & 4392 & 6952 & 9176  & 11808 & 5392 & 8808  & 13048 & 16976 & 5680 & 9856  & 14936 & 20536 \\
     &
      Proposed &
      \textbf{-3936} &
      \textbf{-2936} &
      \textbf{-1960} &
      \textbf{-1040} &
      \textbf{-608} &
      \textbf{1288} &
      \textbf{3320} &
      \textbf{5264} &
      \textbf{752} &
      \textbf{3864} &
      \textbf{6936} &
      \textbf{9936} &
      \textbf{-16} &
      \textbf{4248} &
      \textbf{8280} &
      \textbf{12216} \\ \bottomrule
    \end{tabular}%
    }
    \end{table}

\begin{table}[H]
    \centering
    \caption{FI (bits) of the marked JPEG images under different QFs and different payloads using the proposed method and the previous HS-based RDH methods. The original JPEG images are encoded with the optimized Huffman table.}
    \label{tab:fi_optimized}
    \resizebox{\textwidth}{!}{%
    \begin{tabular}{@{}llrrrrrrrrrrrrrrrr@{}}
    \toprule
    \multirow{2}{*}{\textbf{Image}} & \multirow{2}{*}{\textbf{Method}} & \multicolumn{4}{c}{\textbf{Payload (bits) with QF=30}} & \multicolumn{4}{c}{\textbf{Payload (bits) with QF=50}} & \multicolumn{4}{c}{\textbf{Payload (bits) with QF=70}} & \multicolumn{4}{c}{\textbf{Payload (bits) with QF=90}} \\ \cmidrule(l){3-18} 
     &  & 2000 & 3000 & 4000 & 5000 & 3000 & 5000 & 7000 & 9000 & 4000 & 7000 & 10000 & 13000 & 5000 & 9000 & 13000 & 17000 \\ \midrule
    % Baboon & SP2018 \cite{hou2018reversible} & 2816 & 3928 & 5224 & 6384 & 3264 & 5136 & 7584 & 9400 & 4664 & 8640 & 12344 & 16168 & 7072 & 12168 & 18480 & 23664 \\
    Baboon & TIFS2020 ($\alpha$=0) \cite{he2020reversible}& 2888 & 4072 & 5288 & 6360 & 3080 & 4960 & 7320 & 9832 & 4736 & 8344 & 11968 & 16352 & 7776 & 13512 & 19864 & 24624 \\
     & TIFS2020 ($\alpha$=1) \cite{he2020reversible}& \textbf{1928} & \textbf{2952} & \textbf{4048} & 5376 & \textbf{2344} & \textbf{4024} & \textbf{6480} & \textbf{8072} & \textbf{3328} & \textbf{6376} & \textbf{9888} & 13152 & \textbf{4912} & \textbf{8512} & \textbf{12688} & \textbf{16888} \\
     & SP2020 \cite{li2020reversible}& 2240 & 3528 & 4528 & 5736 & 2600 & 4536 & 6720 & 9208 & 4168 & 7464 & 11152 & 15280 & 7168 & 12160 & 18008 & 21744 \\
     & TCSVT2021 \cite{xiao2020efficient}& 2680 & 3632 & 4832 & 6128 & 2616 & 4656 & 7232 & 9752 & 4328 & 8176 & 11168 & 15552 & 6336 & 12040 & 17880 & 22952 \\
     & Proposed & 2032 & 3176 & 4240 & \textbf{5168} & 3112 & 5072 & 7336 & 8800 & 3616 & 6744 & 9904 & \textbf{12800} & 5432 & 9480 & 13472 & 17680 \\ \midrule
    % Boat & SP2018  \cite{hou2018reversible}& 2552 & 3528 & 4512 & 5584 & 3512 & 5768 & 7768 & 9896 & 4720 & 8384 & 11672 & 14928 & 7840 & 14272 & 19432 & 24112 \\
    Boat & TIFS2020 ($\alpha$=0) \cite{he2020reversible}& 2344 & 3464 & 4456 & 5448 & 3552 & 5824 & 7968 & 10040 & 4800 & 8392 & 11776 & 16112 & 7728 & 13320 & 18376 & 24272 \\
     & TIFS2020 ($\alpha$=1) \cite{he2020reversible}& 2104 & 3304 & 4376 & 5504 & \textbf{2888} & 5256 & 7640 & 9816 & 4048 & 7352 & 10536 & 14184 & 6648 & 11144 & 15552 & 20168 \\
     & SP2020 \cite{li2020reversible}& 2168 & 3192 & 4368 & 5264 & 3192 & 5360 & 7712 & 9840 & 4136 & 7328 & 11016 & 14968 & 6320 & 11112 & 17200 & 21504 \\
     & TCSVT2021 \cite{xiao2020efficient}& 2248 & 3408 & 4288 & 5440 & 3352 & 5704 & 7152 & 9272 & 4568 & 8064 & 11600 & 14816 & 6968 & 12384 & 16872 & 22464 \\
     & Proposed & \textbf{2032} & \textbf{3064} & \textbf{4064} & \textbf{5088} & 3424 & \textbf{4872} & \textbf{6936} & \textbf{8904} & \textbf{3928} & \textbf{6912} & \textbf{10320} & \textbf{13264} & \textbf{5344} & \textbf{9096} & \textbf{13200} & \textbf{17536} \\ \midrule
    % Elaine & SP2018 \cite{hou2018reversible} & 3368 & 4584 & 5504 & 6632 & 4896 & 7624 & 9832 & 11952 & 6560 & 10720 & 14216 & 16936 & 8864 & 14672 & 21176 & 26192 \\
    Elaine & TIFS2020 ($\alpha$=0) \cite{he2020reversible}& 3336 & 4640 & 5472 & 6576 & 4848 & 7440 & 9504 & 11248 & 5896 & 10240 & 13384 & 16040 & 8680 & 14560 & 20000 & 25888 \\
    & TIFS2020 ($\alpha$=1) \cite{he2020reversible}& 2976 & 4352 & 5808 & 7096 & 4360 & 7048 & 9208 & 11456 & 5160 & 8920 & 12376 & 15088 & 6080 & 9880 & 14536 & 18544 \\
    & SP2020 \cite{li2020reversible}& 2832 & 3848 & 4984 & 6400 & 4776 & 6672 & 9504 & 11704 & 5624 & 9144 & 12744 & 15968 & 7448 & 13144 & 18864 & 24648 \\
    & TCSVT2021 \cite{xiao2020efficient}& 3128 & 4488 & 5376 & 6272 & 4792 & 6992 & 9448 & 10888 & 5744 & 9896 & 13048 & 15600 & 7432 & 13200 & 19536 & 25672 \\
    & Proposed & \textbf{2032} & \textbf{2960} & \textbf{4008} & \textbf{5048} & \textbf{3136} & \textbf{5064} & \textbf{7128} & \textbf{9080} & \textbf{4120} & \textbf{6856} & \textbf{10104} & \textbf{12936} & \textbf{4840} & \textbf{9096} & \textbf{13592} & \textbf{17072} \\ \midrule
   % Lena & SP2018 \cite{hou2018reversible} & 2632 & 3952 & 5440 & 6760 & 3728 & 6080 & 8688 & 10520 & 5480 & 9128 & 12672 & 15480 & 7472 & 12776 & 17216 & 21584 \\
   Lena & TIFS2020 ($\alpha$=0) \cite{he2020reversible}& 2576 & 3832 & 5328 & 6760 & 3616 & 5808 & 8768 & 10656 & 5424 & 8944 & 12192 & 15648 & 7680 & 11776 & 16840 & 21168 \\
    & TIFS2020 ($\alpha$=1) \cite{he2020reversible}& 2560 & 4024 & 5312 & 6560 & 3352 & 5672 & 8224 & 10136 & 5152 & 8504 & 11832 & 15144 & 6520 & 10456 & 15576 & 19136 \\
    & SP2020 \cite{li2020reversible}& 2272 & 3768 & 5008 & 6400 & 3144 & 5552 & 8088 & 10616 & 4712 & 7808 & 11232 & 15336 & 6136 & 10416 & 15024 & 19504 \\
    & TCSVT2021 \cite{xiao2020efficient}& 2464 & 3800 & 5216 & 6256 & 3568 & 5520 & 7832 & 10120 & 5136 & 8752 & 11640 & 15184 & 6976 & 10824 & 15584 & 20688 \\
    & Proposed & \textbf{1984} & \textbf{3000} & \textbf{3984} & \textbf{5040} & \textbf{3024} & \textbf{5048} & \textbf{7040} & \textbf{9064} & \textbf{3880} & \textbf{6872} & \textbf{10096} & \textbf{13024} & \textbf{5032} & \textbf{9072} & \textbf{13272} & \textbf{17416} \\ \bottomrule
    \end{tabular}%
    }
    \end{table}

\par
\begin{figure}[H]
    \centering
    \subfigure[QF=30]{
        \includegraphics[width= 0.4\textwidth]{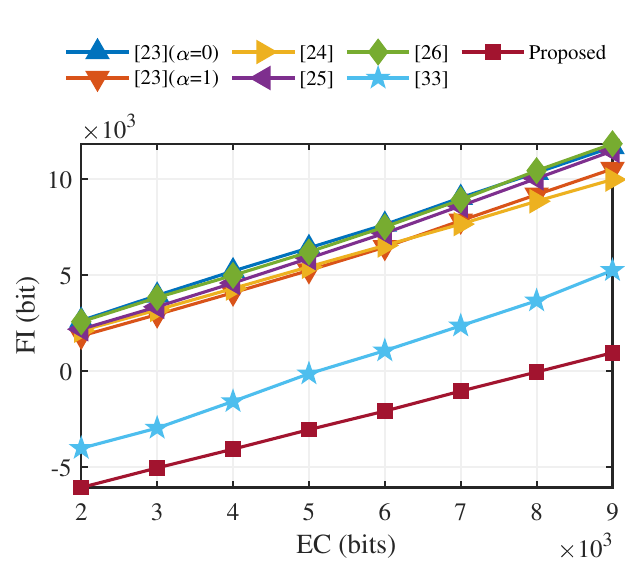}}
    \subfigure[QF=50]{
        \includegraphics[width= 0.4\textwidth]{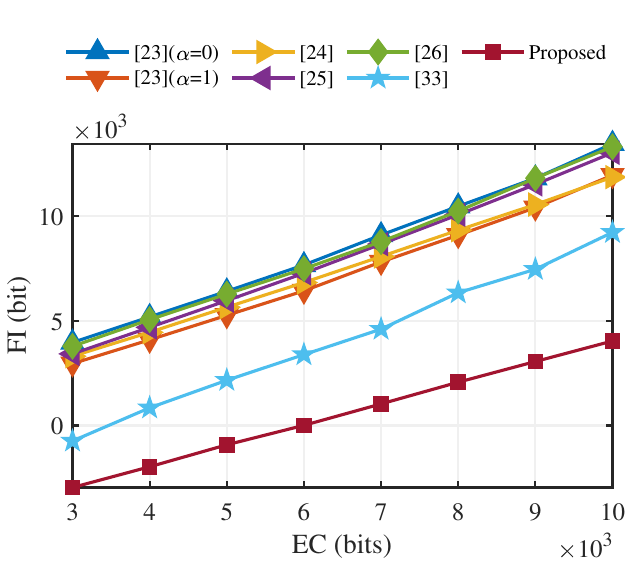}}

    \subfigure[QF=70]{
        \includegraphics[width= 0.4\textwidth]{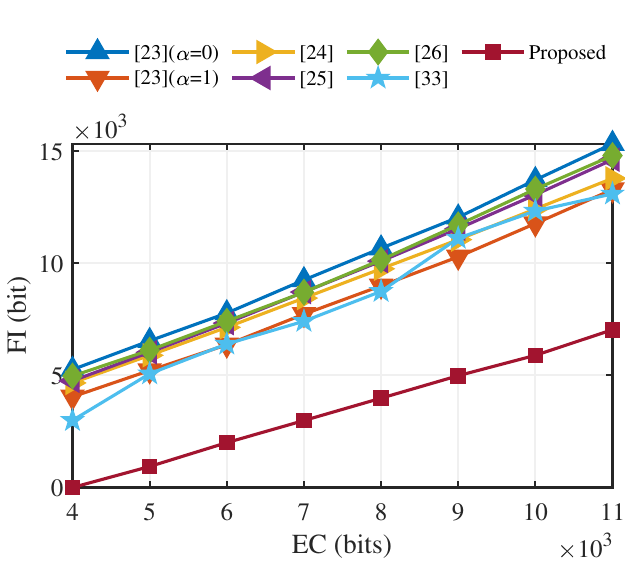}}
    \subfigure[QF=90]{
        \includegraphics[width= 0.4\textwidth]{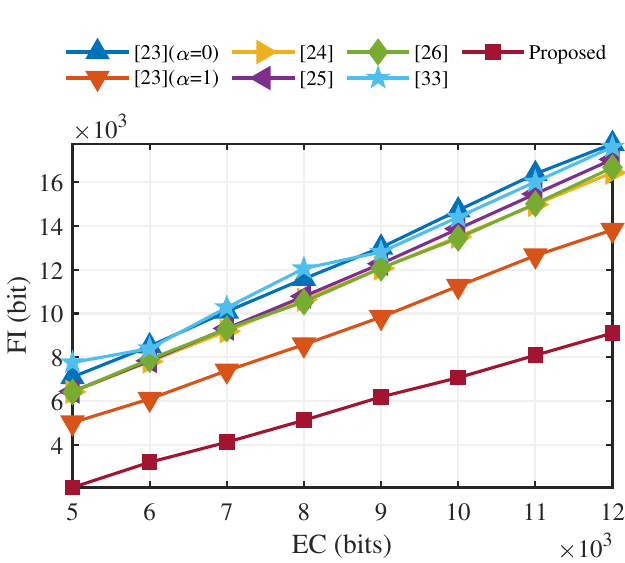}}
    \caption{Average FI of 200 random images from the BOSSbase image database. The original JPEG images are encoded with the standard Huffman table.}
    \label{fig_d_fi}
\end{figure}

\begin{figure}[H]
    \centering
    \subfigure[QF=30]{
        \includegraphics[width= 0.4\textwidth]{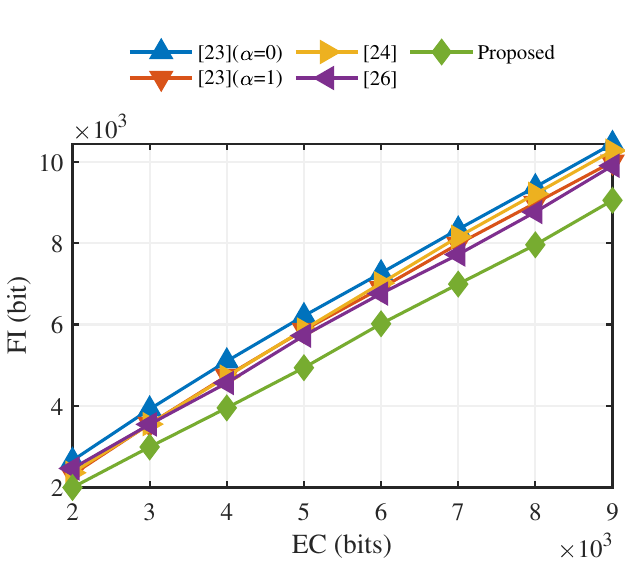}}
    \subfigure[QF=50]{
        \includegraphics[width= 0.4\textwidth]{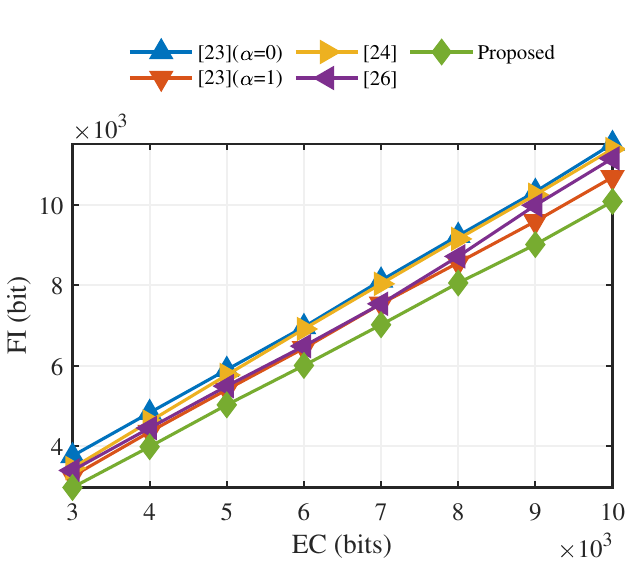}}

    \subfigure[QF=70]{
        \includegraphics[width= 0.4\textwidth]{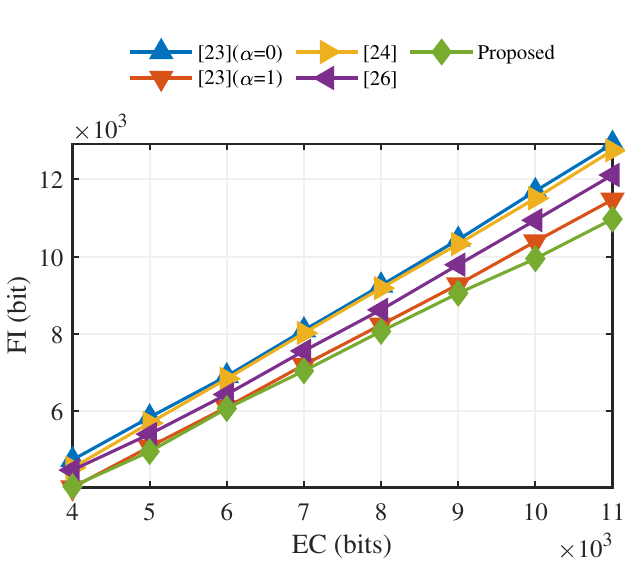}}
    \subfigure[QF=90]{
        \includegraphics[width= 0.4\textwidth]{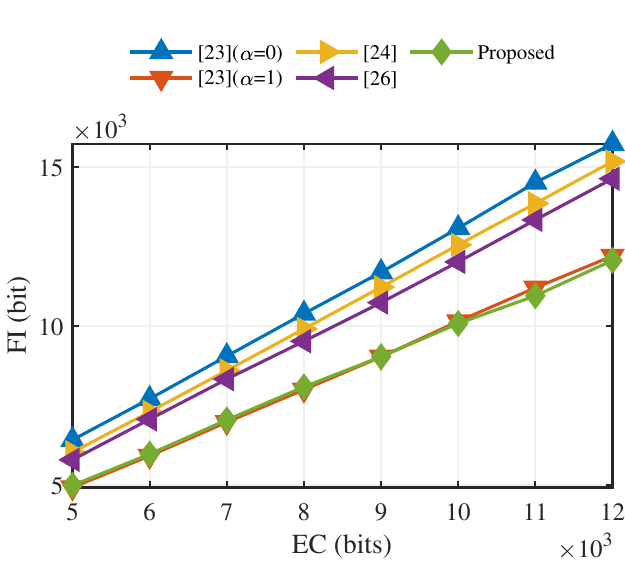}}
    \caption{Average FI of 200 random images from the BOSSbase image database. The original JPEG images are encoded with the optimized Huffman table.}
    \label{fig_o_fi}
\end{figure}

 We also tested the file size preservation performance on BOSSbase databases. Two hundred random images from the BOSSbase image database were embedded with additional data of different payloads. The average FIs of both the std-bitstream and the opt-bitstream are shown in Fig. \ref{fig_d_fi} and Fig. \ref{fig_o_fi}, respectively. It can be observed in Fig. \ref{fig_d_fi} and Fig. \ref{fig_o_fi} that the average FI caused by the proposed method is significantly smaller than that of the previous RDH methods at different payloads and QFs.

\subsection{Evaluation on image quality}\label{section6.3}

\par
\begin{figure}[H]
    \centering
    \subfigure[QF=30]{
        \includegraphics[width= 0.4\textwidth]{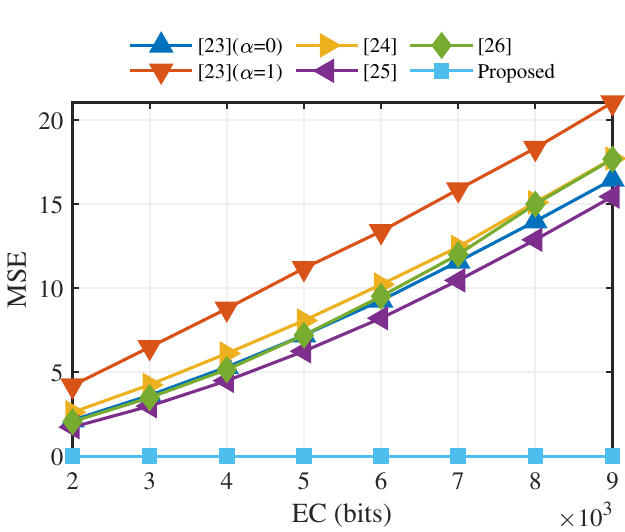}}
    \subfigure[QF=50]{
        \includegraphics[width= 0.4\textwidth]{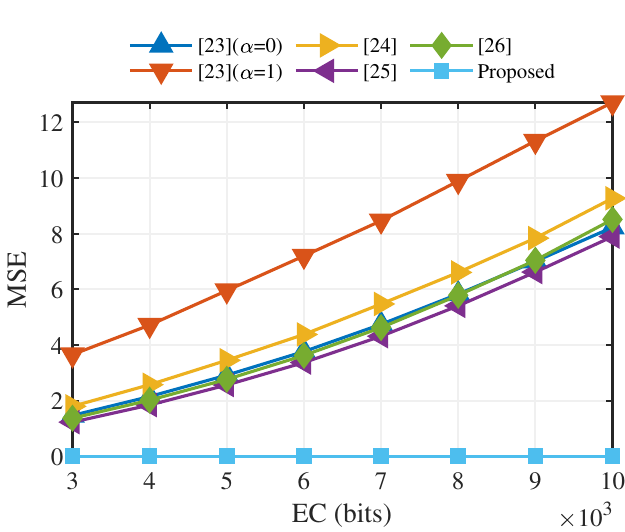}}

    \subfigure[QF=70]{
        \includegraphics[width= 0.4\textwidth]{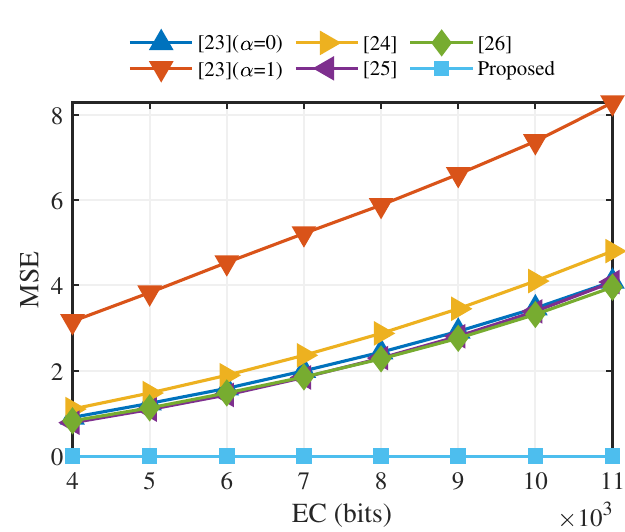}}
    \subfigure[QF=90]{
        \includegraphics[width= 0.4\textwidth]{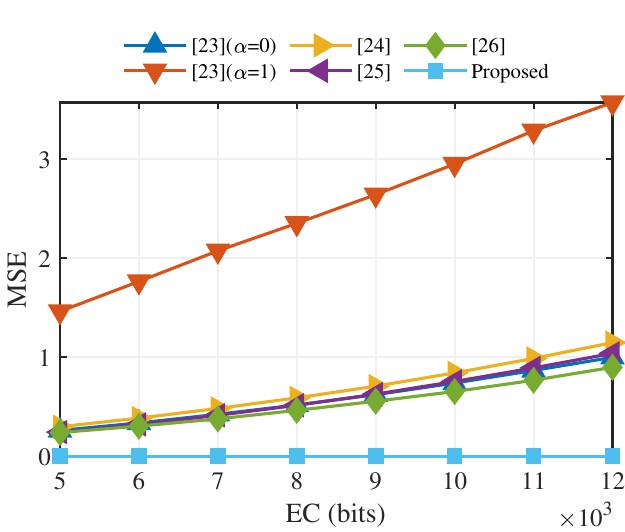}}
    \caption{Average MSE of 200 images from the BOSSbase image database. The original JPEG images are encoded with the optimized Huffman table.}
    \label{fig_mse}
\end{figure}

 To compare the performance on image quality of marked JPEG images, we tested the MSE of the 200 images from the BOSSbase image database. The results are shown in Fig. \ref{fig_mse}. As shown in Fig. \ref{fig_mse}, the MSEs of the proposed method are always equal to zero, which means the marked JPEG images generated by the proposed method maintain an unchanged visual quality compared to the original JPEG images.

\subsection{Evaluation on complexity}\label{section6.4}

\par The running times of four test images are listed in Table. \ref{tab:time_cm}, where the text in bold indicates the shortest running time of the methods in the same case. It can be observed that the running times of the method \cite{du2020gvm} are the shortest in most cases. The method \cite{zhang2020reversible} is the most time-consuming. Our proposed method is competitive among these high-capacity CM-based methods in terms of runtime. 

\begin{table}[H]
    \centering
    \caption{Running time (s) of test images with different QFs using the proposed method and two state-of-the-art CM-based methods \cite{du2020gvm, qiu2020optimized,zhang2020reversible}. The original JPEG images are encoded with the standard Huffman table.}
    \label{tab:time_cm}
    \resizebox{\textwidth}{!}{%
    \begin{tabular}{@{}llrrrrrrrrr@{}}
    \toprule
    \textbf{Image} & \textbf{Method} & \textbf{QF=10} & \textbf{QF=20} & \textbf{QF=30} & \textbf{QF=40} & \textbf{QF=50} & \textbf{QF=60} & \textbf{QF=70} & \textbf{QF=80} & \textbf{QF=90} \\ \midrule
    Baboon & TCSVT2021-I \cite{qiu2020optimized} & 1.734 & 1.972 & 3.423 & 3.495 & 4.281 & 5.804 & 5.558 & 5.604 & 6.299 \\
     & Proposed & \textbf{0.522} & \textbf{0.722} & \textbf{0.942} & \textbf{1.023} & \textbf{1.205} & \textbf{1.299} & \textbf{1.458} & \textbf{1.713} & \textbf{2.514} \\ \cmidrule(l){2-11} 
     & TCSVT2021-II \cite{qiu2020optimized} & 3.192 & 2.461 & 3.679 & 3.779 & 4.705 & 6.147 & 6.048 & 6.082 & 7.122 \\
     & TDSC2020 \cite{du2020gvm} & \textbf{0.438} & \textbf{0.669} & 0.922 & 1.057 & 1.232 & 1.308 & 1.624 & 1.943 & 2.645 \\
     & Proposed & 0.543 & 0.708 & \textbf{0.892} & \textbf{1.026} & \textbf{1.207} & \textbf{1.301} & \textbf{1.467} & \textbf{1.833} & \textbf{2.505} \\ \cmidrule(l){2-11} 
     & JVCIR2020 \cite{zhang2020reversible} & 169.987 & 220.959 & 290.210 & 501.693 & 544.242 & 618.708 & 707.202 & 458.541 & 690.555 \\
     & Proposed & \textbf{0.515} & \textbf{0.721} & \textbf{0.853} & \textbf{0.971} & \textbf{1.079} & \textbf{1.192} & \textbf{1.332} & \textbf{1.629} & \textbf{2.226} \\ \midrule
    Boat & TCSVT2021-I \cite{qiu2020optimized} & 1.378 & 2.014 & 2.487 & 4.396 & 5.474 & 5.248 & 5.441 & 6.506 & 7.039 \\
     & Proposed & \textbf{0.418} & \textbf{0.503} & \textbf{0.688} & \textbf{0.667} & \textbf{0.728} & \textbf{0.808} & \textbf{0.899} & \textbf{1.09} & \textbf{1.748} \\ \cmidrule(l){2-11} 
     & TCSVT2021-II \cite{qiu2020optimized} & 1.461 & 2.015 & 2.717 & 4.481 & 6.074 & 5.573 & 6.105 & 7.674 & 7.717 \\
     & TDSC2020 \cite{du2020gvm} & \textbf{0.293} & \textbf{0.449} & \textbf{0.539} & \textbf{0.605} & \textbf{0.684} & 0.938 & 0.946 & \textbf{1.204} & 1.894 \\
     & Proposed & 0.412 & 0.600 & 0.589 & 0.693 & 0.725 & \textbf{0.813} & 0.958 & 1.221 & \textbf{1.703} \\ \cmidrule(l){2-11} 
     & JVCIR2020 \cite{zhang2020reversible} & 80.529 & 185.825 & 297.026 & 421.591 & 517.990 & 666.214 & 254.186 & 483.803 & 916.252 \\
     & Proposed & \textbf{0.399} & \textbf{0.526} & \textbf{0.582} & \textbf{0.652} & \textbf{0.725} & \textbf{0.800} & \textbf{0.905} & \textbf{1.066} & \textbf{1.542} \\ \midrule
    Elaine & TCSVT2021-I \cite{qiu2020optimized} & 1.313 & 1.499 & 1.749 & 1.758 & 1.767 & 1.863 & 2.377 & 2.351 & 3.804 \\
     & Proposed & \textbf{0.385} & \textbf{0.440} & \textbf{0.510} & \textbf{0.577} & \textbf{0.659} & \textbf{0.819} & \textbf{0.784} & \textbf{1.117} & \textbf{1.761} \\ \cmidrule(l){2-11} 
     & TCSVT2021-II \cite{qiu2020optimized} & 1.184 & 1.809 & 1.979 & 1.934 & 1.919 & 2.009 & 2.453 & 2.572 & 4.069 \\
     & TDSC2020 \cite{du2020gvm} & \textbf{0.239} & \textbf{0.332} & \textbf{0.412} & \textbf{0.475} & \textbf{0.632} & \textbf{0.676} & 0.834 & \textbf{1.081} & 2.063 \\
     & Proposed & 0.367 & 0.521 & 0.530 & 0.587 & 0.708 & 0.833 & \textbf{0.820} & 1.179 & \textbf{1.782} \\ \cmidrule(l){2-11} 
     & JVCIR2020 \cite{zhang2020reversible} & 23.740 & 50.431 & 82.316 & 115.965 & 168.235 & 180.639 & 218.327 & 294.595 & 451.699 \\
     & Proposed & \textbf{0.344} & \textbf{0.429} & \textbf{0.487} & \textbf{0.542} & \textbf{0.617} & \textbf{0.698} & \textbf{0.798} & \textbf{1.013} & \textbf{1.616} \\ \midrule
    Lena & TCSVT2021-II \cite{qiu2020optimized} & 1.235 & 1.546 & 1.654 & 2.070 & 2.216 & 2.471 & 2.953 & 3.891 & 6.088 \\
     & Proposed & \textbf{0.358} & \textbf{0.425} & \textbf{0.487} & \textbf{0.533} & \textbf{0.636} & \textbf{0.662} & \textbf{0.776} & \textbf{0.886} & \textbf{1.251} \\ \cmidrule(l){2-11} 
     & TCSVT2021-II \cite{qiu2020optimized} & 1.555 & 1.863 & 1.871 & 2.116 & 2.47 & 2.513 & 3.209 & 4.723 & 6.541 \\
     & TDSC2020 \cite{du2020gvm} & \textbf{0.255} & \textbf{0.335} & \textbf{0.469} & 0.596 & \textbf{0.541} & \textbf{0.611} & \textbf{0.718} & \textbf{0.943} & 1.435 \\
     & Proposed & 0.371 & 0.444 & 0.488 & \textbf{0.577} & 0.628 & 0.665 & 0.847 & 0.949 & \textbf{1.349} \\ \cmidrule(l){2-11} 
     & JVCIR2020 \cite{zhang2020reversible} & 67.357 & 118.483 & 140.226 & 240.800 & 204.149 & 177.214 & 222.365 & 400.918 & 636.610 \\
     & Proposed & \textbf{0.362} & \textbf{0.438} & \textbf{0.493} & \textbf{0.548} & \textbf{0.605} & \textbf{0.647} & \textbf{0.730} & \textbf{0.880} & \textbf{1.225} \\ \bottomrule
    \end{tabular}%
    }
\end{table}

\par
\begin{figure}[H]
        \centering
        \subfigure[QF=30]{
            \includegraphics[width= 0.4\textwidth]{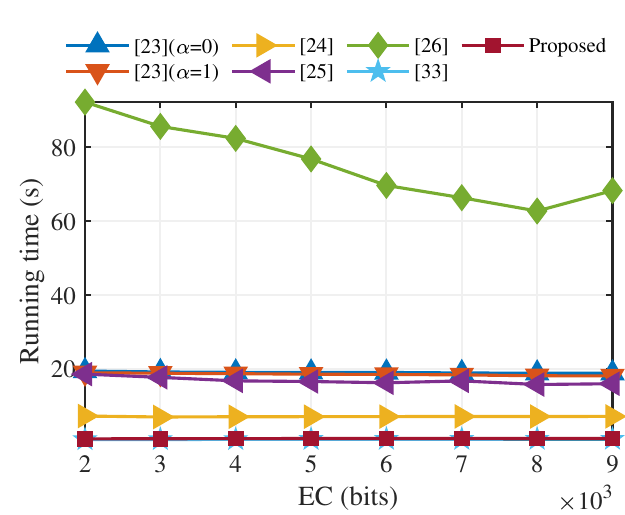}}
        \subfigure[QF=50]{
            \includegraphics[width= 0.4\textwidth]{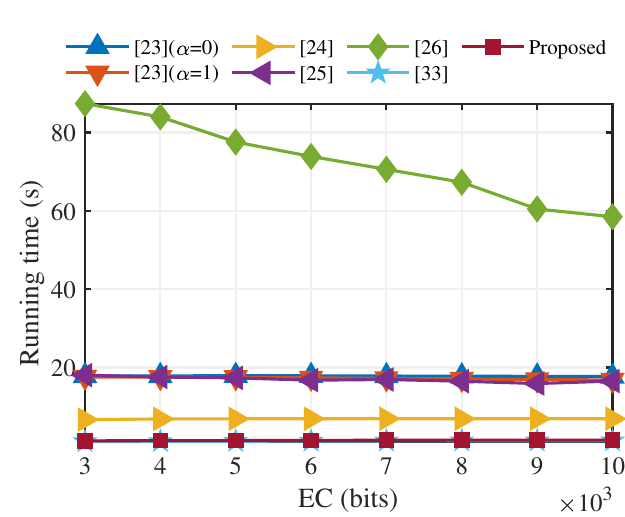}}
    
        \subfigure[QF=70]{
            \includegraphics[width= 0.4\textwidth]{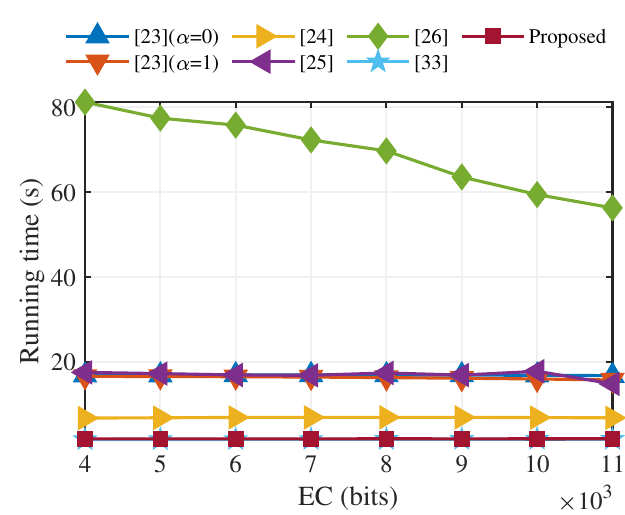}}
        \subfigure[QF=90]{
            \includegraphics[width= 0.4\textwidth]{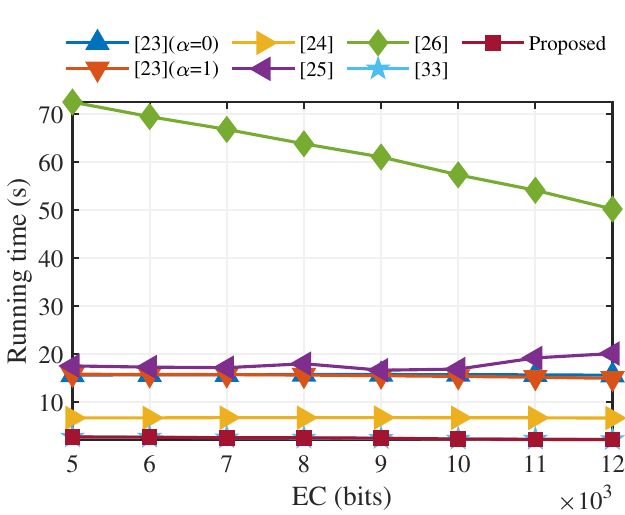}}
        \caption{Average running time of 200 images from the BOSSbase image database. The original JPEG images are encoded with the standard Huffman table.}
        \label{fig_d_time}
\end{figure}

 The average running time of 200 images from the BOSSbase images database is illustrated in Fig. \ref{fig_d_time}. The original JPEG images in Fig. \ref{fig_d_time} are encoded with the standard Huffman table. As shown in Fig. \ref{fig_d_time}, the running time of the proposed method is faster than the previous HS-based method but a little slower than the CM-based RDH method proposed by Du $et~al.$ \cite{du2020gvm}. This is because finding a nearly optimal solution using GA is time-consuming. 

\section{Conclusion}\label{section7}

\par In this study, we proposed a new CM-based RDH framework to improve the performance of file size preservation and applicability. Compared to the traditional framework of CM-based RDH, the proposed framework adjusts the code-mapping construction step and adds a Huffman table customization step. Using the proposed framework, a new CM-based RDH method can be obtained by designing a specific optimization algorithm. To demonstrate this, we proposed a new CM-based RDH method using a GA to obtain the optimal code-mapping relationship. The experimental results demonstrated that our method achieves impressive file size preservation, visual quality, and time complexity performance. In the future, we will continue to explore the following directions.

\begin{enumerate}
    \item In our proposed method, solving the optimization algorithm is the key issue. Therefore, more efficient optimization algorithms can be investigated. 
    \item Recently, reversible data hiding in encrypted images has attracted a considerable attention. The performance of CM-based RDH for encrypted JPEG bitstream should therefore also be investigated.  
\end{enumerate}

\section*{Acknowledgment}
\par This research work is partly supported by National Natural Science Foundation of China (61872003, 62172001).

\bibliography{mybibfile}

\end{document}